\documentclass[twocolumn,preprint]{aastex62} 
\usepackage[utf8x]{inputenc}
\usepackage{amsmath,amsfonts,amssymb,mathrsfs,bm,enumerate,graphicx}
\usepackage[T1]{fontenc}
\usepackage[all]{hypcap}
\usepackage{ulem} 

\usepackage{ae,aecompl}
\hypersetup{breaklinks,colorlinks,citecolor=blue,linkcolor=magenta}

\newcommand{\pderiv}[1]{\frac{\partial}{\partial #1}}
\newcommand{\ppderiv}[2]{\frac{\partial #1}{\partial #2}}
\newcommand{\deriv}[1]{\frac{d}{d #1}}

\newcommand{\del}{\mathbf{\nabla}}
\renewcommand{\div}{\del \cdot}
\newcommand{\avg}[1]{\left\langle #1 \right\rangle}
\renewcommand{\vec}[1]{\mathbf{#1}}

\newcommand{\be}{\begin{eqnarray}}
\newcommand{\ee}{\end{eqnarray}}

\shorttitle{Pileups around Planets in Disks}
\shortauthors{Dempsey, Lee, \& Lithwick}

\begin{document}

\title{Pileups and Migration Rates for Planets in Low Mass Disks}
\author[0000-0001-8291-2625]{Adam M. Dempsey}
\author[0000-0002-5319-3673]{Wing-Kit Lee}
\author{Yoram Lithwick}
\affiliation{Center for Interdisciplinary Exploration and Research in Astrophysics (CIERA)
and
Department of Physics and Astronomy
Northwestern University \\
2145 Sheridan Road
Evanston, IL 60208
USA}

\correspondingauthor{Adam M. Dempsey}
\email{adamdempsey2012@u.northwestern.edu}

\begin{abstract}
We investigate how planets interact with viscous accretion disks, in the limit that the disk is sufficiently low mass that the planet migrates more slowly than the disk material. 
In that case, the disk's surface density profile is determined by the disk being in viscous steady state (VSS), while overflowing the planet's orbit. 
We compute the VSS profiles with 2D hydrodynamical  simulations, and show that disk material piles up behind the planet, with the planet effectively acting as a leaky dam.  
Previous 2D hydrodynamical simulations missed the pileup effect because of incorrect boundary conditions, while previous 1D models greatly overpredicted the pileup due to the neglect of non-local deposition.
Our simulations quantify the magnitude of the pileup for a variety of planet masses and disk viscosities.  
We also calculate theoretically the magnitude of the pileup for moderately deep gaps, showing good agreement with simulations.  
For very deep gaps, current theory is inadequate, and we show why and what must be understood better.
The pileup is important for two reasons. 
First, it is observable in directly imaged protoplanetary disks, and hence can be used to diagnose the mass of a planet that causes it or the viscosity within the disk. 
And second,  it determines the planet's migration rate.  
 Our simulations determine a new Type-II migration rate (valid for low mass disks), and show how it connects continuously with the well-verified Type-I rate.

\end{abstract}


\section{Introduction}

Protoplanetary disks are being observed in ever-increasing detail, e.g., via imaging and
spectral studies \citep{2011ARA&A..49...67W,2014prpl.conf..497E,2015ApJ...808L...3A,2016ApJ...820L..40A,2018ApJ...869L..41A}.
The inferred properties of these disks can be used to test theories for
protoplanetary disk evolution and planet formation. 
For example, many imaged disks exhibit bands and gaps that
may be sculpted by  planetary mass companions
\citep[e.g.,][and others]{2015ApJ...808L...3A,2015ApJ...806L..15K,2016A&A...595A.114D,2016PhRvL.117y1101I,2017A&A...600A..72F,2018ApJ...866..110D,2018ApJ...869L..47Z}. 
In addition,  SEDs (and images) of so-called transitional disks reveal that they
  have inner holes, which might be emptied out by planets 
\citep{2011ApJ...729...47Z,2011ApJ...732...42A,2011ApJ...738..131D,2012ApJ...755....6Z,2014prpl.conf..497E,2016A&A...585A..58V,2019NatAs.tmp..329H}.

The planet-disk interaction problem has been studied extensively
\citep[for reviews see e.g.,][]{1993prpl.conf..749L,2012ARA&A..50..211K,2014prpl.conf..667B}.  
Planets torque material in the disk by launching spiral waves, which then damp and deposit their
angular momentum in the disk.  
 If the planet is sufficiently massive, the torques
  open up a gap in the disk \citep{1986ApJ...307..395L}. 
An inevitable corollary of the planet torquing the disk is the disk torquing the planet, and the 
resulting migration of the planet. 
Historically, the migration timescales were thought to fall into two main categories:
Type I for
low mass planets that are unable to open gaps in their disks; and Type II for
 high mass planets  that open nearly infinitely deep gaps, locking them
into the disk's viscously driven accretion
\citep{1997Icar..126..261W}.

 Simplified 1D models have been constructed for the mutual evolution
of the planet and disk when the gap is very deep \citep[e.g.,][]{1995MNRAS.277..758S,1999MNRAS.307...79I,1997Icar..126..261W,2010PhRvD..82l3011L,2012MNRAS.427.2660K,2012MNRAS.427.2680K}.  
But, as we shall show in this paper, such models have a serious difficulty due to the fact that they usually assume (either explicitly or implicitly) ``local deposition'': i.e., that waves damp immediately after being
launched.
In reality, however, deposition is non-local, as waves transport angular momentum
from where they are excited to where they are damped.
Under local deposition, gaps become extremely deep---with the depth depending exponentially on the planet mass \citep[e.g.,][]{2007ApJ...667..557T,2010PhRvD..82l3011L,2014ApJ...782...88F,2015MNRAS.448..994K}. 
In early work \citep{1995MNRAS.277..758S,1999MNRAS.307...79I,1997Icar..126..261W}, it was assumed that massive enough planets opened infinitely
deep gaps, and thus no material could flow across the planet's orbit.  
In  \citet{2012MNRAS.427.2660K,2012MNRAS.427.2680K}, 
exponentially deep gaps were considered. 
They made the same low-disk-mass approximation  that we make in this paper:
that the disk remains in viscous steady state as the planet migrates.
However, their work is based on the local approximation.
We discuss their work further
in \S\ref{sec:kocsis}.
A key result of many of the aforementioned 1D models
 \citep[e.g.,][]{1995MNRAS.277..758S,2010PhRvD..82l3011L,2012MNRAS.427.2660K,2012MNRAS.427.2680K} is that they produce an enhancement of gas exterior to the planet's orbit (termed a pileup), and  this pileup follows the planet as it migrates inwards.

More recently, there has been a significant amount of progress in understanding deep gaps around
large planets.
Based initially on hydrodynamical simulations, several authors \citep[e.g.,][]{Crida:2006vh,2014ApJ...782...88F} found that even very massive planets do not open exponentially clean gaps. 
Instead, their results follow a (non-exponential) scaling relationship which can be derived analytically if one assumes most of the angular momentum injected by the planet comes from nearby the planet \citep{2015MNRAS.448..994K,2015ApJ...807L..11D}.
 \citet{2017ApJ...835..146D}, \citet{2017PASJ...69...97K}, and \citet{2019arXiv190611256D}
 expanded the parameter space covered by these simulations and have found similar scaling relations -- even showing that gaps in 3D are similar to gaps in 2D \citep{2016ApJ...832..105F}.
However, none of the hydrodynamical studies have reproduced the pileup effect seen in the local deposition 1D models.
That seems somewhat puzzling as one would expect that a very massive planet should act as a barrier to accreting material, and even if the barrier is partial, it should slow down the accretion of the gas.
This might be important observationally, as the pileup may be responsible for the inner hole
in transitional disks (or, more precisely, the pileup is the observed part of the disk with the inner hole).

A second outstanding question for planets that open deep gaps---in addition to the existence of a pileup---is what is their migration
rate?
This has been addressed recently by several authors. 
\citet{2014ApJ...792L..10D} and \citet{2015A&A...574A..52D} found that gap opening planets are not locked into the disk's viscous evolution, but instead migrate at a range of rates set primarily by the disk-to-planet mass ratio, with larger disk masses resulting in faster migration.
\citet{2018ApJ...861..140K} found similar results and provided an empirical formula for the migration rate which smoothly connects the non-gap opening regime to the deep gap regime. 
In slight tension with these results, however, the simulations of \citet{2018A&A...617A..98R} showed that the deep gap migration rates, while not being exactly the Type II rate, were still proportional to the disk's viscosity. 

In the present work, we address both of these questions, focusing on a
 particularly simple case: when the disk is sufficiently low-mass that the planet
migration rate is slower than the disk material. As we shall see, this results in a particularly clean setup, 
since the disk's viscous steady state structure can be studied while ignoring planet migration.

The outline of the paper is as follows. 
In \S\ref{sec:1d} we set up the planet-disk interaction problem in low mass disks, and study it analytically 
to the extent that we can.  Our main result is that  there is a single
quantity that needs to be determined: the total amount of angular momentum put into the disk by the
planet ($\Delta T$) when the disk is in viscous steady state. This quantity controls both the pileup and the planet's migration rate.  In \S\ref{sec:simulations}--\ref{sec:results}
 we turn to hydrodynamical simulations with the 
primary goal of determining $\Delta T$:
in \S\ref{sec:simulations} we outline our numerical method, focusing on our  new boundary conditions which allow the disk to settle into the steady-state solution described in  \S \ref{sec:1d}; and
in \S \ref{sec:results} we present
the results of a suite hydrodynamical simulations.
In \S\ref{sec:discussion}, we consider some implications of our simulations, including the planet migration
rate.
Finally, we summarize and list some open questions in \S\ref{sec:conclusions} and \S\ref{sec:open}.

\section{Planets in Low-mass disks} \label{sec:1d}

Our basic assumption throughout this paper is that the accretion disk is sufficiently 
low mass that the planet migrates more slowly than the disk material.
As we show in \S\ref{sec:mig}, for the parameter-range  that we consider,
 a disk qualifies as low-mass if it is slightly less massive than the planet; in fact, in some cases the disk can even be more massive than the planet and still qualify as low-mass.
 Such disks are  relevant for Jupiter-mass planets, and may also be relevant
 for terrestrial planets during the late stages of planet formation.
We shall also assume that the planet is circular with radius $r_p$ and mean motion $\Omega_p$.

Our primary goal is to calculate the planet-disk torque ($\Delta T)$\footnote{We denote it $\Delta T$ because it is a sum of positive torque on exterior material and negative torque on interior material.} once the disk has reached viscous steady state (VSS). 
This torque is important for two reasons. 
First, it affects the surface density profile of the disk in VSS, leading to a pileup of material outside of the planet's orbit.
Second, it forces the planet to migrate by removing the planet's angular momentum. 
As the planet migrates, the disk passes through successive VSS solutions. 
Therefore in the aforementioned low-disk-mass limit, we may obtain the planet's migration rate by considering the dynamics on timescales long enough for the disk to reach VSS, while neglecting dynamics on the longer migration time.

Three  dimensionless parameters   affect $\Delta T$ in a non-trivial way: the planet-star mass ratio ($q$), the strength of  viscosity (e.g., as parameterized by the \citeauthor{1973A&A....24..337S} $\alpha$), and  the aspect ratio of the disk ($h\equiv H/R$). 
A fourth potential parameter is the accretion rate of the disk in VSS ($\dot{M}$), or equivalently the overall amplitude of the surface density. 
But with the fairly standard assumptions that we shall make, $\Delta T\propto \dot{M}$
\footnote{In particular, the proportionality $\Delta T\propto \dot{M}$ relies on the assumption
that the disk is locally isothermal.
We suspect that using a more realistic equation of state will not change our
results significantly, but leave the verification to future work.}
, i.e., the dependence on $\dot{M}$ is trivial. As a result,
we shall calculate
 the dimensionless torque $\Delta T/(\dot{M}\ell_p)$ where $\ell_p=r_p^2\Omega_p$ is the specific angular momentum of the planet, and this  will be a function of  three  dimensionless parameters ($q,\alpha,h$). 

\subsection{Excitation and Deposition of Angular Momentum}
\label{sec:exdep}

With the planet's orbit fixed and circular, there are two timescales on which the disk's properties evolve---the wave and viscous timescales. 
On the faster  wave timescale,  the planet excites waves in the disk, and  these propagate away from the planet where they damp by viscosity or shocks. 
On this timescale wave steady state (WSS) is reached, meaning that the wave pattern becomes stationary in the rotating reference frame of the planet.
On the viscous timescale, the azimuthally-averaged (``mean'') surface density reacts to the damping of the waves, and VSS is reached. 
The wave timescale is  $\sim {\rm( orbital\ time)}/h$, because the group velocity of pressure waves is of order the sound speed \citep{Ogilvie:2002fu}.
The viscous timescale is $\sim {\rm (orbital\ time)}/(\alpha h^2)$, and  therefore  significantly longer.

Angular momentum is transferred from the planet to the disk in a two-stage process: 
(i) Excitation: the planet excites waves at Lindblad resonances, where it transfers angular momentum to the waves;
and (ii) Deposition: after the waves propagate, they deposit their angular momentum in the disk.\footnote{
For clarity, we ignore here  the complication that
angular momentum is also transferred to co-orbital material, which does not launch propagating waves.  We show below that co-orbital torques are typically sub-dominant.
 }
The distinction between excitation and deposition is sometimes ignored in the literature 
\citep[e.g.,][although the former reference discusses some of the effects of this distinction]{1997Icar..126..261W,2010PhRvD..82l3011L,2012MNRAS.427.2660K,2012MNRAS.427.2680K}.
Nonetheless,
 it is of crucial importance
\citep[e.g.,][]{1982Icar...52...14L,1983Icar...53..207G,2001ApJ...552..793G,2002ApJ...569..997R,2002ApJ...572..566R,2010ApJ...724..448M,2015ApJ...807L..11D,2015MNRAS.448..994K,2017PASJ...69...97K,2018MNRAS.479.1986G}.
In particular, whereas excitation is straightforward to calculate from linear theory
 \citep{1980ApJ...241..425G}, 
it is deposition that controls the surface density profile of the disk.

We  make the above discussion quantitative via equations that track angular momentum transfer. 
The reader uninterested in technical details may skip the remainder of this subsection without great loss.
We consider a 2D disk, in which the dynamical variables are $\{\Sigma,v_r,v_\phi\}$ in standard notation. 
Where convenient, we shall also employ $\ell \equiv r v_\phi$ (specific angular momentum) and $\Omega\equiv v_\phi/r$ as surrogates for $v_\phi$. 
Variables are decomposed into ``mean'' and ``wave'' components, denoted by brackets and primes respectively, e.g., $\Sigma= \langle \Sigma\rangle+\Sigma'$, where $\langle\Sigma\rangle\equiv \oint \Sigma d\phi/(2\pi)$.

In Appendix \ref{sec:app_deriv}, we start  from the general 2D equations of motion for a disk with shear viscosity (Eqs.\ \eqref{eq:app_2d_sig}-\eqref{eq:app_2d_mom}) to derive exact
equations for three angular momentum densities: 
the total ($\propto\avg{\Sigma\ell}$), wave ($\propto \avg{\Sigma'\ell'}$), and mean flow ($\propto\avg{\Sigma}\avg{\ell}$), with the following results. 
\begin{enumerate}
\item Total:
    \be \label{eq:avl}
        \pderiv{t}2\pi r \avg{ \Sigma \ell } + \pderiv{r} \left(2\pi r\avg{ \Sigma v_r \ell} +F_\nu \right)= {t_{\rm ex}} ,
    \ee 
    where the excitation torque density\footnote{
   We adopt the convention of representing torque densities (i.e., torque per unit radius) by lower case $t$, and torques (or angular momentum fluxes) by capitalized $F$ or $T$. 
    } is 
    \be
     \label{eq:exex}
        t_{\rm ex} = - 2\pi r\avg{\Sigma' \ppderiv{\Phi'}{\phi} } \ ,
    \ee
    in which $\Phi'$ is the wave component of the planet's potential, and
    the viscous torque is
    \be
    \label{eq:fnux}
        F_\nu =  -2\pi r^2\avg{ \nu \Sigma \left( r\ppderiv{\Omega}{r} + \frac{1}{r} \ppderiv{v_r}{\phi} \right) }  .
    \ee
   
    An approximate form for $t_{\rm ex}$ is derived by  \cite{1980ApJ...241..425G}, resulting in the well-known ``standard torque
    formula'' ($t_{\rm ex}\propto \Sigma q^2/|r-r_p|^4$).
    We shall show from our simulations that, with minor modifications, the standard
    torque formula is  of satisfactory accuracy, even when the waves are nonlinear. 

 For $F_\nu$, we may typically neglect its non-wave contribution to approximate
\be \label{eq:fnuapprox}
    F_\nu \approx -2\pi r^3 \nu \avg{  \Sigma} \partial_r\avg{\Omega} , \label{eq:fnuapprox1}
\ee   as we shall verify in our hydrodynamical simulations   (see  Appendix \ref{sec:app_approx} and also \citet{2017PASJ...69...97K}).  A further approximation is to assume that $\langle\Omega\rangle$
is nearly Keplerian, as is typically true everywhere except near the bottom of a deep gap,
resulting 
 in the familiar form
 \be
    F_\nu\approx 3\pi \nu\langle \Sigma\rangle\langle \ell\rangle  \label{eq:fnuapprox2} .
 \ee
    
     Equation \eqref{eq:avl} shows that the disk locally conserves angular momentum, aside from that input by the planet ($t_{\rm ex}$).  
The total torque that the planet applies to the disk is the  quantity that we ultimately desire:
\be
    \Delta T=\int_0^\infty t_{\rm ex}dr .
\ee

\item Wave:
    \be \label{eq:lwave}
      \pderiv{t} 2 \pi r \avg{\Sigma' \ell'} + \pderiv{r} F_{\rm wave} =t_{\rm ex}  - t_{\rm dep} ,
    \ee
    where the flux of angular momentum carried by the waves is
    \begin{align}
    F_{\rm wave} &= 2 \pi r^2 \avg{\Sigma v_r v_\phi' } \label{eq:fwavex}
    \\
  &=  2 \pi r^2\left( \avg{\Sigma} \avg{ v_r' v_\phi'} + \avg{ \Sigma'v_r' v_\phi'}+\avg{v_r}\avg{\Sigma'v_\phi'} \right) . \label{eq:fwavex3}
    \end{align}
    In the latter expression, the first term $\propto \langle v_r'v_\phi'\rangle$ is
    the usual wave flux that is conserved in the linearized adiabatic problem 
    without viscosity \citep[e.g.,][]{1979ApJ...233..857G}; 
    the second is a triple correlation that can become of comparable importance when the waves are nearly nonlinear;
and the third is generally negligible outside of the co-orbital zone because $\avg{v_r}$ is ${\cal O}(\alpha h^2)$ (for an example see Appendix \ref{sec:app_approx}).
   
   The quantity $t_{\rm  dep}$ in Eq.\ (\ref{eq:lwave}) is the deposition torque density;  $t_{\rm dep}$ is displayed explicitly in Eq.\ \eqref{eq:tdep_def}, but for present purposes it suffices to note
    that it  vanishes wherever there are no waves.\footnote{Our expression for $t_{\rm dep}$ does not vanish when the viscosity  vanishes. 
    That is a consequence of using a locally isothermal equation of state.  If the more realistic
    locally adiabatic equation of state were used, 
    our expression for
    $t_{\rm dep}$ would vanish at zero viscosity
    \citep{2019ApJ...878L...9M}. 
    Of course, the case of zero viscosity is not  physical: at small viscosity, 
    dissipation occurs at shocks \citep{2001ApJ...552..793G}.
     }

The wave equation may typically be simplified: since the wave timescale is  shorter than the viscous one, when considering viscous evolution one may  drop the $\partial/\partial t$ in that equation, yielding the wave steady state (WSS) equation
\be \label{eq:wss}
  \deriv{r}F_{\rm wave} \approx t_{\rm ex}  - t_{\rm dep} \quad {\rm (WSS)} .
\ee
To appreciate the implication, let us focus for definiteness on orbital radii $r>r_{\rm p}$, in which case the outer torque excited by the planet is $T_+=\int_{r_p}^\infty t_{\rm ex}dr=\int_{r_p}^\infty t_{\rm dep}dr$, where the latter relation follows from the fact that the wave flux vanishes at infinity and at the planet (ignoring co-orbital torques). 
In other words, the total exterior torque {\it excited} by the planet is equal to that {\it deposited} into the mean flow, with $F_{\rm wave}$ the intermediary that transports angular momentum from where it is excited (Lindblad resonances) to where it is deposited.

\item Mean flow:
    \be \label{eq:lavg2}
        \pderiv{t}  2 \pi r \avg{\Sigma}\avg{\ell} + \pderiv{r} \left(- \dot{M} \avg{\ell}  +F_\nu\right) = t_{\rm dep}  ,
     \label{eq:lav}
    \ee
    where
    \be
        \dot{M}= - 2\pi r \langle\Sigma v_r\rangle ,
    \label{eq:mdot}
    \ee
    is the mass accretion rate\footnote{Our sign convention is such that $\dot{M} > 0$ corresponds to inwards mass accretion, i.e. $v_r < 0$.}.

\end{enumerate}
Equations \eqref{eq:avl}--\eqref{eq:mdot} follow from the general 2D equations without
approximation (aside from those denoted explicitly with $\approx$ that we have made
for simplicity)---in particular, they do not assume that wave quantities are smaller than mean ones, or a specific form for the shear viscosity or equation of state. 
Note that Eq.\ \eqref{eq:avl} is equal to the sum of Eqs.\ \eqref{eq:lwave} and \eqref{eq:lav} because $\langle \Sigma \ell \rangle=\langle \Sigma'\ell'\rangle+\langle\Sigma\rangle\langle\ell\rangle$. 

\citet{2017PASJ...69...97K} perform a similar decomposition
to that presented above, 
but for the steady-state equations. As a result, they do not distinguish  between WSS and
VSS.
Equation (\ref{eq:lav}) is well-known, e.g., \citet{1997Icar..126..261W} and \citet{2002ApJ...572..566R}. But the decomposition into waves and mean flow allows for an explicit general expression for $t_{\rm dep}$, given in Eq.\ \eqref{eq:tdep_def}.

\subsection{Viscous Evolution}

The mean flow equation (Eq.\ \ref{eq:lav}) governs the evolution of the mean surface density on the viscous timescale.
It is equivalent to the standard angular momentum equation for a planet-less viscous accretion disk \citep[e.g.,][]{1974MNRAS.168..603L}, aside from the term $t_{\rm dep}$, which is caused by the transfer of angular momentum from waves to the mean flow as they damp. 
The fluxes in this equation are the viscous $F_\nu$, as in Eq.\ \eqref{eq:avl}, and $-\dot{M}\avg{\ell}$, which represents the inward advective transport of the mean flow's $\avg{\ell}$\footnote{ Note that  $\dot{M}\propto \avg{\Sigma v_r}=\avg{\Sigma}\avg{v_r}+\avg{\Sigma'v_r'}$, and therefore the waves participate in the advection. 
If one wished, the  $\avg{\Sigma'v_r'}$ term could be transferred into the definition of $t_{\rm dep}$, while at the same time adding it into  $F_{\rm wave}$. 
We choose not to do so because we wish the viscous evolution equations (Eqs.\ \eqref{eq:lav} and \eqref{eq:mcons}) to form a closed set for the quantities $\dot{M}$ and $\avg{\Sigma}$, once $t_{\rm dep}$ is known.}.

As the disk evolves towards VSS, its viscous evolution is determined by Eq.\ \eqref{eq:lav}, together with mass conservation:
\be \label{eq:mcons}
    \pderiv{t} 2\pi r\avg{\Sigma}=\pderiv{r} \dot{M} ,
\ee
where $\dot{M}$ is defined in Eq.\ \eqref{eq:mdot}.
Equations \eqref{eq:lav} and \eqref{eq:mcons} form a closed set of equations for
$\avg{\Sigma}$ and $\dot{M}$ because
(i) $\avg{\ell}$ is nearly Keplerian, aside from a (typically small) correction which
may be determined from $\avg{\Sigma}$ by radial pressure balance;
(ii)  $F_\nu$ may be approximated by Eq.\ \ref{eq:fnuapprox1};
and (iii) the $t_{\rm dep}$ profile can be calculated  from the $\langle \Sigma\rangle$ profile, because the waves may be considered to be in WSS.
However, determining the $t_{\rm dep}$ profile theoretically is difficult, even under the WSS assumption, and we shall resort to numerical simulations to determine it (\S\ref{sec:simulations}-\ref{sec:results})\footnote{\citet{2001ApJ...552..793G}, and \citet{2002ApJ...569..997R}
calculate  $t_{\rm dep}$ for sub-thermal-mass planets ($q \lesssim h^3$)
 and  in the absence of viscosity; see also
 \citet{2015ApJ...807L..11D} and
  \citet{2018MNRAS.479.1986G}
  who use a somewhat crude approximation to $t_{\rm dep}$ in order to determine self-consistent gap profiles.
But those results
   are not directly applicable to the higher-mass planets and viscous
 disks that we consider in this paper.}.

\subsection{Viscous Steady State (VSS)}

In VSS, Eqs.\ \eqref{eq:lav} and \eqref{eq:mcons} imply
\begin{align}
\dot{M} &= {\rm const} ,\\
{d\over dr}\left( -\dot{M}\avg{\ell}+F_\nu \right) &= t_{\rm dep} \label{eq:tdep}  \ .
\end{align}
For a given $t_{\rm dep}$ profile, the solution
 of the second equation is trivially
\be  \label{eq:fnustar}
F_\nu(r)= \dot{M}\avg{\ell}+F_* + \int_{r_{i}}^r t_{dep}(r')dr' \ .
\ee
Here, $F_*$ is an integration constant and $r_{i}$ is arbitrary, but we shall choose it to be sufficiently inwards of the planet's orbit that the waves have damped by then, and so $t_{dep}$ vanishes at $r<r_{i}$\footnote{We ignore here the possibility that
waves can sometimes reach the inner edge of the disk if the viscosity is small enough \citep{2002ApJ...569..997R}.}.
The constant $F_*$ represents the angular momentum flux injected at the inner edge of the disk, e.g., by the star.
It is often chosen to yield $F_\nu = \Sigma=0$ near the star's surface \citep{1973A&A....24..337S,1974MNRAS.168..603L}.
But because the first term in Eq.\ \eqref{eq:fnustar} increases as $\avg{\ell} \propto r^{1/2}$,  we may discard the constant $F_*$ more than a few stellar radii away from the star. 
The VSS solution may therefore be written as
\be
 \label{eq:fnu1}
F_\nu(r)= \dot{M}\avg{\ell}+ \int_{r_{i}}^r t_{dep}(r')dr'  {\ \ \ \rm (VSS)} \ .
\ee
provided $r$ is far enough from the star (as we assume to be true henceforth). 
We shall make use of this VSS solution extensively in our analysis below. 
Note that the profile of $F_\nu(r)$ immediately determines
 the $\avg{\Sigma}$ profile  after inserting an approximate form for $F_\nu$  (Eq.\ \eqref{eq:fnuapprox} or \eqref{eq:fnuapprox2}). In other words, it is $t_{\rm dep}$
(rather than $t_{\rm ex}$)
that directly controls the
surface density profile of the disk.

\subsubsection{VSS solution far from the planet: connecting $\Delta T$ to the pileup and the migration rate}

\begin{figure}
    \centering

	\includegraphics[trim={0.3cm 0.3cm 0.22cm 0},clip,width=0.48\textwidth]{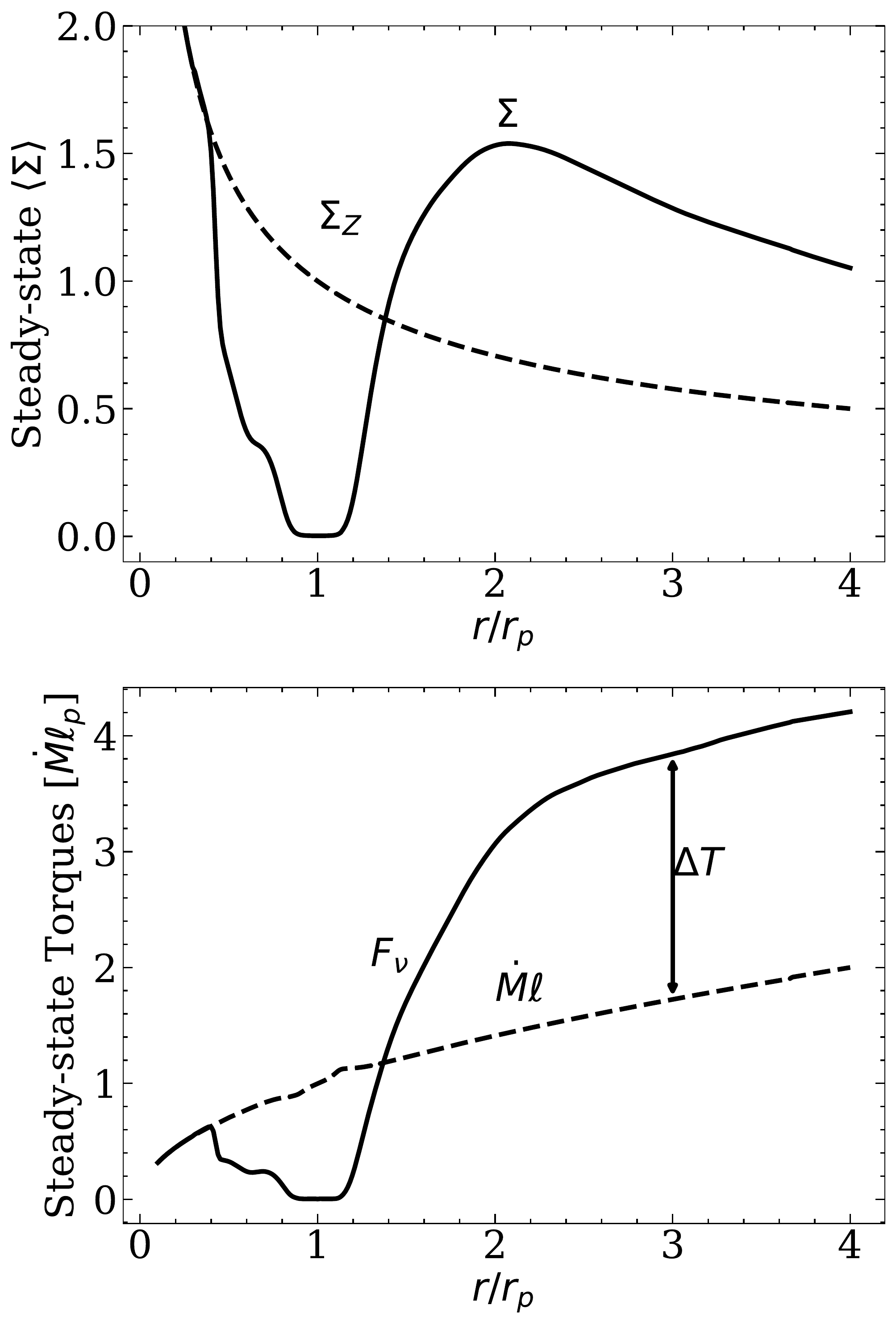}
    \caption{Illustration of the VSS solution given in Eqs.\ \eqref{eq:fnu1} and \eqref{eq:sigma_sol}. 
    The top panel shows an example steady-state $\Sigma$ profile compared to the ZAM $\Sigma$ profile given in Eq.\ \eqref{eq:ZAM}. 
    The bottom panel shows the corresponding $F_\nu$ profile compared to the ZAM $F_\nu$ profile, $\dot{M} \ell$. The constant offset between $F_\nu$ and $\dot{M} \ell$ at large radii corresponds to the total torque input by the planet, $\Delta T$. }
    \label{fig:schematic}
\end{figure}

We
 may apply  the above solution to determine  the density profile far from the planet:
\be
  \label{eq:fnusol}
    F_\nu= 	\dot{M}\ell\times 
    \begin{cases}
        1 &  { \ }  r < r_{i}  \\
         1+{\Delta T\over \dot{M}{\ell}}  & { \ } r>r_{o} 
    \end{cases}
\ee
where $r_{o}$ is the  distance beyond which $t_{dep}\approx 0$ and $\Delta T$ is the total deposited torque, which must be equal to the excited torque (Eq.\ \ref{eq:wss}). We have
dropped angled brackets, because the waves are damped in this domain.
Using $F_\nu\approx 3\pi\nu\Sigma \ell$
 (Eq.\ \ref{eq:fnuapprox2}), the VSS surface density is 
\be \label{eq:sigma_sol} 
\Sigma &=& {\dot{M}\over 3\pi\nu}\times
	\begin{cases}
		1 &  { \ }  r < r_{i}  \\
		1+{\Delta T\over \dot{M}\ell}  & {\ } r>r_{o} 
	\end{cases}
\ee  
The quantity 
\be \label{eq:ZAM}
    \Sigma_Z \equiv {\dot{M}\over 3\pi\nu} ,
\ee
that appears in Eq.\ \eqref{eq:sigma_sol} is the well-known solution for a planet-less accretion disk far from the star \citep{1974MNRAS.168..603L}---or equivalently one with $F_*=0$. 
For ease of reference below, we call it the ``zero-angular-momentum-flux'' (or ZAM) solution, whence the subscript Z.

Steady-state solutions given by Eqs.\ \eqref{eq:fnusol} and \eqref{eq:sigma_sol} with $\Delta T \neq 0$ have previously been studied in the context of the disk inner boundary where $\Delta T$ is the torque of the central star on the disk \citep[e.g.,][]{1973A&A....24..337S,1974MNRAS.168..603L}, and in the context of circumbinary disks where $\Delta T$ is the total torque of the binary on the disk \citep[e.g.,][]{1995MNRAS.277..758S,2012MNRAS.427.2660K,2012MNRAS.427.2680K,2013ApJ...774..144R,2016ApJ...827..111R,2017MNRAS.466.1170M,2017MNRAS.469.4258T,2019ApJ...871...84M}. 

To better understand  the VSS solution described by Eqs.\ \eqref{eq:fnusol} and \eqref{eq:sigma_sol} we plot an illustrative example in Figure \ref{fig:schematic}. 
The curves are taken from one of our hydrodynamical simulations described in \S\ref{sec:results}. 
The top panel shows that far inside of the planet, $\Sigma=\Sigma_Z$, while 
far outside there is a pileup relative to $\Sigma_Z$.
The bottom panel shows that far outside the planet $F_\nu-\dot{M}\ell=\Delta T$, which is 
spatially constant; the value of $\Delta T$ determines the height of the pileup in $\Sigma$.

The torque deposited into the disk, $\Delta T$, comes at the expense of the planet's orbit, implying that the planet's instantaneous migration rate is
    $\dot{r}_p = -2 r_p \Delta T/(M_p \ell_p)$
where $M_p$ is the planet mass and where we assume that the planet maintains a circular orbit and does not accrete any material. 
A consequence of the above  is that whenever $\Delta T>0$ the planet will migrate inwards, and will be accompanied by a pileup outside of its orbit.

\subsubsection{Calculation of $\Delta T$ in moderately deep gaps, and the difficulty
with very deep gaps} \label{sec:dt_theory}
\label{sec:moderate}

We apply here the VSS equation to determine
$\Delta T$ for gaps that are 
 ``moderately deep'' (to be defined shortly).
The results will be shown to match  those from simulations
 for gaps that are $\gtrsim 25\%$ of the background density.
Although moderately deep gaps are only of moderate interest---particularly if one 
is interested in large pileups---we present the theory here because  it helps clarify
the results of the simulations to be presented shortly, and  it also shows  why the 
theory is much more difficult for  deeper gaps. 
The theory for moderately deep gaps was developed by
\citet{2015ApJ...807L..11D} and \citet{2015MNRAS.448..994K}.
We mostly follow their approach, but extend it to calculate the two-sided torque ($\Delta T$).

The standard torque formula 
is $t_{\rm ex}\propto \Sigma q^2/|r-r_p|^4$, with a cutoff at
$|r-r_p|\lesssim h$
\citep{1980ApJ...241..425G}\footnote{We ignore co-orbital 
torques in this section, but consider their impact in \S\ref{sec:torques} in the context of our
simulations. }
.
Therefore, provided that the gap is not too deep, 
most of the excited torque comes from a distance $\sim h$ from the planet. 
For such ``moderately deep'' gaps, one may set the inner excited torque 
to $T_-\equiv \int_0^{r_p} t_{\rm ex}dr={\rm const}\times q^2\Sigma_p/h^3$, where
$\Sigma_p$ is the value at the planet.\footnote{We implicitly assume that $\Sigma$ does not vary significantly between
$r_p$ and $r_p-h$, which is expected to be true even when there is a gap,
because the lengthscale of the gap is set by $t_{\rm dep}$ rather than $t_{\rm ex}$.
Our simulation results  support this expectation (see Figure \ref{fig:kall} below).
} A similar argument applies to the outer
excited torque ($T_+$), but with a different constant.  In order to obtain these constants,
one must 
 account for the detailed shape of $t_{\rm ex}$ near the torque cutoff, which we do by
 numerically solving the linear equations of motion; details are in  Appendix \ref{sec:app_linear}.\footnote{To isolate 
 the one-sided Lindblad torques we make use of the property that the linear wave flux far from the planet carries all of the Lindblad torque. 
We solve the linear equations on top of a background  $\Sigma = \Sigma_Z$, with a small enough viscosity such that we can measure the wave angular momentum flux far from the planet \citep{1993Icar..102..150K}.   Since the wave flux is not conserved in the locally isothermal equations \citep[see e.g.,][]{Lee:2016ku,2019ApJ...878L...9M}, we also assume $c_s = h = {\rm const}$ for the linear calculation. The effect of the $c_s$ gradient should be sub-dominant as the important Lindblad resonances are located $\sim h$ away from the planet.}
We find
\be 
T_\pm&\approx&C_\pm {q^2\over h^3}\Sigma_p r_p^4\Omega_p^2 , \label{eq:Tpm}\\
C_+&\approx&0.48 \label{eq:cp} , \\
C_-&\approx&-0.36 \label{eq:cm} ,
\ee
where these $C_\pm$ are applicable for $h=0.05$ and $\Sigma_Z\propto r^{-1/2}$, which are the values we
shall use in our simulations. More general expressions can be found in, e.g., \citet{2002ApJ...565.1257T}.

We may now obtain  $\Sigma_p$ from the VSS equation
(Eq.\ \ref{eq:fnu1}) at $r_p$:
\be
\left(F_\nu-\dot{M}\ell\right)_{r_p} &=& \int_0^{r_p} t_{\rm dep}(r')dr'   \label{eq:far_inner}  \\
&=& T_- , \label{eq:fint}
\ee
where the second equality follows from the excited torque being equal to the deposited torque
(Eq.\ \ref{eq:wss}).
 Setting $F_\nu=3\pi\nu\Sigma \ell$, 
 $\dot{M}=3\pi\nu\Sigma_Z$, and $\nu=\alpha h^2 \ell $ we obtain
 for the depth of the gap
 \be \label{eq:sm1}
{\Sigma_p\over \Sigma_{Z,p}} \approx
{1\over 1+ |C_-| K/(3\pi)}
\approx \frac{1}{1 + 0.04 K}  \ ,
\label{eq:sigsig}
 \ee
where
 \be 
 K \equiv {q^2\over \alpha h^5} ,
 \ee
  is a commonly used parameter  that measures the relative strength of a planet's gravitational torque
  ($\propto q^2/h^3$ at distance $h$)
   to the disk's viscous torque   ($\propto \alpha h^2$ at distance $h$) \citep[][]{1997Icar..126..261W,2015ApJ...807L..11D,2015MNRAS.448..994K,2017PASJ...69...97K}.
 Inserting  $\Sigma_p$ into Eq.\ (\ref{eq:Tpm}) yields the one-sided torques
 $T_+$ and $T_-$. The two-sided  torque $\Delta T=T_++T_-$ is then
\be \label{eq:dt1}
\Delta T &\approx&
{C_++C_-\over 3\pi}{K\over 1+0.04K}\dot{M}\ell_p \\
&\approx&
\frac{0.013 K}{1 + 0.04 K} \dot{M} \ell_p \label{eq:dt1n}  \ .
\ee
We see that the reason for the existence of a non-vanishing $\Delta T$ is that the
exterior torques exceed the interior torques ($|C_+|>|C_-|$) by ${\cal O}(h)$, 
as
 is well-known from studies of Type I migration \citep{1980ApJ...241..425G,1997Icar..126..261W}.
\citet{2015ApJ...807L..11D} and \citet{2015MNRAS.448..994K} previously derived Eq.\ \eqref{eq:sigsig}, while  \citet{2018ApJ...861..140K} derived Eq.\ \eqref{eq:dt1n} with the numerical coefficient extracted from their simulations.

We show below that this first-principles prediction for gap depth and $\Delta T$ agrees well with simulation results for moderately deep gaps.\footnote{We have assumed in our derivation that the torque cutoff occurs at $h$, which is true
for sub-thermal-mass planets.  We have not explored the case of super-thermal-mass planets
because in our simulations the planets that produce moderately deep gaps are mostly sub-thermal.}
At first glance, it might appear surprising that one may predict the gap depth and torques
without any knowledge of the $t_{\rm dep}$ profile. The reason is that once one knows {\it where}
the torques are excited, one may calculate the ratio of excited torque to surface density at that
location.  
Since the total excited torque is equal to the total deposited torque, and since the deposited torque
determines the surface density, one then has a closed system of equations. We may  
illustrate this by equating the timescale for a planet to open a gap with that required 
for viscosity to close it, as must be true in VSS.
 The latter time is $t_{\rm close}\sim x_{\rm gap}^2/\nu$, where $x_{\rm gap}$
is the width of the gap, which is set by the width of the $t_{\rm dep}$ profile.  The former is
$t_{\rm open}\sim L/T_-$, where $L$ is the angular momentum required to vacate
 material from the
gap, $L\sim (\Sigma_{Z,p}-\Sigma_p)x_{\rm gap}^2$ (setting
$r_p=\Omega_p=1$ here for simplicity).  Equating the two timescales, we see that $x_{\rm gap}^2$
cancels, and with $T_-\sim \Sigma_p q^2/h^3$ and $\nu\sim \alpha h^2$, we find $\Sigma_p/\Sigma_{Z,p}\sim 1/(1+K)$, in agreement with the form of Eq.\ (\ref{eq:sigsig}); a slightly more
careful calculation can nearly reproduce the order-unity coefficient multiplying $K$.

The above discussion suggests that constructing a theory for very deep gaps will be  difficult. 
Once a gap is sufficiently deep,  most of the torque will be excited beyond  $h$ from the planet \citep{2012ApJ...758...33P,2018MNRAS.479.1986G}.   
In order to determine that distance, one needs to know the amount of torque deposited 
inside of that distance, since that will affect the surface density 
there, which in turn controls where the torque is excited.
But, as previously emphasized, understanding deposition is difficult.
An additional, though related, difficulty is that in our  derivation for moderately deep gaps we assumed that
$\Sigma$ was nearly constant between the inner and outer excitation locations ($r_p-h$ and $r_p+h$). But for very deep gaps that is no longer true, and the jump in $\Sigma$ from its inner
to its outer excitation location is also determined  by the   deposition of torque within that zone.
Further discussion of very deep gaps, in light of our simulation results, will be presented in \S\ref{sec:model}.

\section{Numerical Method} \label{sec:simulations}

Our main goal in the remainder of the paper is to calculate $\Delta T$ with hydrodynamical
simulations, and to understand and explain the results theoretically. 
Henceforth, we shall set the planet's orbital radius and 
mean motion to $r_p=1$ and $\Omega_p=1$, which sets the length and time
units. Note that one could also set $\dot{M}$ to unity, because the viscous
evolution equations are linear in $\dot{M}$ for the locally isothermal equation of
state that we adopt.
However, we prefer to keep dependences on $\dot{M}$ explicit to avoid potential confusion (where helpful, we also keep some dependences on $r_p$ and $\Omega_p$ explicit).

Our numerical setup is mostly standard, with one main exception: 
 the inner and outer
boundary conditions are based on the VSS solution far from the planet
(Eq.\ \ref{eq:sigma_sol}), which allows us to find a pileup where others 
have not.

We use the GPU-accelerated FARGO3D code \citep{2016ApJS..223...11B} to evolve the 2D hydrodynamical equations of motion, Eqs.\ \eqref{eq:app_2d_sig}-\eqref{eq:app_2d_mom}. 
We take the equation of state to be locally isothermal, $P = c_s^2 \Sigma  = h^2 (r\Omega_K)^2\Sigma$, where $\Omega_K$ is the Keplerian orbital frequency and $h$ is the  aspect ratio, which we fix at  $h=0.05$. 
Viscosity is modelled explicitly, with kinematic shear viscosity $\nu=\alpha c_s^2/\Omega_K=\alpha h^2 r^2\Omega_K$ and constant $\alpha$.

The planet  is modeled as a softened  gravitational potential with softening length  $\epsilon = 0.6 h= 0.03$.
This value  approximates  the vertically-averaged 3D potential to within $\sim 10\%$ for distances $|r-1| > h $ \citep{Muller:2012hx}.
We present further details of our numerical setup in Appendix \ref{sec:app_num}. 

\subsection{Boundary Conditions} \label{sec:bcs}
Boundary conditions are implemented in FARGO3D 
with layers of ghost cells interior to our inner boundary, located at $r_i$, and exterior to our outer boundary, located at $r_o$.
At the outer boundary, we wish to supply the system with a steady mass accretion rate $\dot{M}$ without injecting any angular momentum at the inner boundary, i.e., we want $F_*=0$ in Eq.\ \eqref{eq:fnu1}.
At the inner boundary the solution should match onto the ZAM solution (Eq.\ \ref{eq:ZAM}), where the $\dot{M}$ that appears in the solution should be $\dot{M}|_{r_{i}}$ (rather than the injected $\dot{M}|_{r_{o}}$).
To account for this, we match onto the ZAM solution by ensuring that $\nu\Sigma$ is constant across the boundary, i.e., we set the value of $\Sigma$ in ghost cells at $r<r_{i}$ such that $\nu\Sigma$ in those ghost cells is constant and equal to the value in the cell at $r=r_{i}$.
For $v_r$ in the inner ghost cells, we set it to the ZAM expression $v_r=3 \nu /(2 r)$, i.e. the value consistent with $\Sigma$ and $\dot{M}|_{r_i}$. Finally, for $v_\phi$, we set it to its (pressure-supported) Keplerian value by 
extrapolating from the cell at $r=r_{i}$. 
We also ensure that the flow is axisymmetric at $r_{i}$ by adopting a wave-killing zone  between $r_{i}<r<r_{i,wkz}$, where we damp $v_r$ to its azimuthal average (see Eq.\ \ref{eq:wkz}). 
By damping only $v_r$ and not $v_\phi$ or $\Sigma$, we ensure that the wave-killing procedure conserves angular momentum (and mass), and therefore all of the angular momentum excited by the planet---and carried by density waves---is deposited into the disk within the computational domain.

At the outer boundary, we seek to match onto the exterior VSS solution $\Sigma=\Sigma_Z(1+ \Delta T/(\dot{M} \ell))$ (Eq.\ \ref{eq:sigma_sol}), where $\dot{M}$ is the mass to be injected at $r_{o}$,
and $\Delta T$ is unknown beyond the fact that it should be a constant number in VSS. 
One way to avoid the difficulty of not knowing $\Delta T$ beforehand is to extend $r_{o}$ to a sufficiently large value that $\Delta T\ll \dot{M}\ell$, in which case the exterior VSS solution is $\Sigma\approx \Sigma_Z$ \citep[as was done in e.g.,][]{2017MNRAS.466.1170M,2019ApJ...871...84M}. 
But rather than making the computational domain so large, we note that in the exterior VSS solution $d F_\nu/dr=\dot{M}d\ell/dr$ (Eq.\ \ref{eq:tdep} with $t_{\rm dep} =0$), which provides  a condition on the gradient of $\Sigma$ at $r_{o}$---given an input value for $\dot{M}$.
To apply this condition, we set in the ghost cells (at $r>r_{o}$) $F_\nu=F_\nu\vert_{r_{o}}+(\ell-\ell\vert_{r_{o}})\dot{M}$, i.e., we set $\Sigma$ in the ghost cells according to the relation $3\pi\nu\Sigma\ell=(3\pi\nu\Sigma\ell)\vert_{r_{o}}+(\ell-\ell\vert_{r_{o}})\dot{M}$.
To set the value of $v_r$ in the ghost cells, we then use $v_r=-\dot{M}/(2\pi r \Sigma)$, and for $v_\phi$ we set it to the pressure-corrected and extrapolated Keplerian value, as for the inner disk. 
As was the case for the inner boundary, we enforce a wave-killing-zone near the outer boundary between $r_{o,wkz} < r < r_o$.

We note that \citet{2017MNRAS.466.1170M}  used, for a subset of their circumbinary disk simulations,
 an outer boundary condition  similar in 
spirit to ours:
they fixed $\Sigma$ at the outer boundary to the value given by Eq.\ \eqref{eq:sigma_sol}, where the $\Delta T$ was measured from a previous iteration of the simulation.

\subsection{Iterative approach to VSS}

In order to reach VSS we must integrate the equations of motion for several viscous times at the outer boundary, which is prohibitively long.
For $\alpha \sim 10^{-4}$ an outer viscous time is $\sim 5$ million planet orbits. To run that long
on, for example, one K80 GPU takes a wallclock time of over a year, at our typical timestep of 0.003 planet orbits.
Therefore, for our small $\alpha$ simulations we adopt an iterative approach. At each
iterative step,
 we start from 
a profile for $\langle\Sigma(r)\rangle$. We  then run a FARGO3D simulation to WSS, which is
much shorter than VSS (\S \ref{sec:exdep}). The result of that simulation determines $t_{\rm dep}$, which 
in turn determines $\langle\Sigma(r)\rangle$ from the VSS equation (Eq.\ \ref{eq:fnu1}), and hence
can be used to initiate the next iterative step.

We consider a simulation to have reached VSS when the time-averaged $\dot{M}$ throughout the domain is
within 10\% of the forced $\dot{M}$ at the outer boundary. In our simulations with $q\gtrsim 2\times 10^{-3}$ the disk becomes eccentric \citep{2006MNRAS.368.1123G,2006A&A...447..369K,2008A&A...487..671K,2014ApJ...782...88F,2017MNRAS.467.4577T}. 
The disk eccentricity makes defining a steady-state difficult as there is a long precession timescale in the disk and it is uncertain if the disk eccentricity should persist in steady-state. 
For these reasons we simply omit these simulations from our analysis, leaving a more
detailed study of such disks to future work. 

\subsection{Resolution}

In all of our simulations, the computational domain extends from $(r_i, r_o) = (0.3, 3.68)$ with uniform spacing in azimuth and $\ln(r)$.
The edges of the wave-killing zones are at $(r_{i,wkz}, r_{o,wkz})=(0.46,3.0)$. 
The number of grid cells in each dimension is $N_\phi \times N_r = 1005 \times 401$, which provides near square cells (i.e. $\Delta r \sim r \Delta \phi$) and corresponds to  eight cells per scale-height. 
To check convergence, we have also run each simulation at the cruder resolution $N_\phi \times N_r = 502 \times 200$.
We find that the low resolution total torque agrees with the high resolution result to within $\sim 10\%$, on average, and $\sim 30\%$ in the worst case (see Table \ref{tab:sims}).

\section{Numerical Results} \label{sec:results}

\begin{figure}
    \centering
	\includegraphics[trim={0.4cm 0.4cm 0.4cm 0},clip,width=0.48\textwidth]{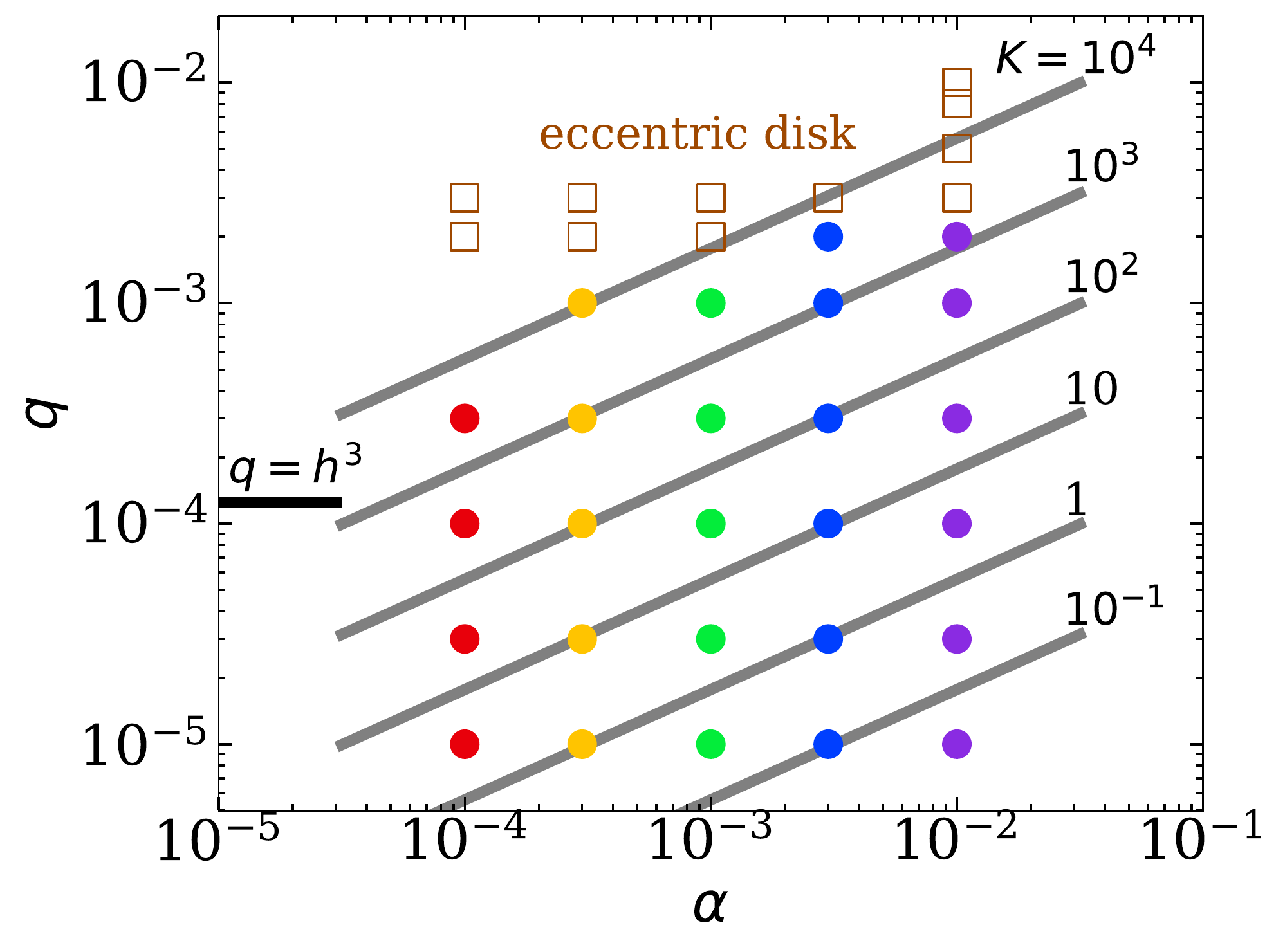}
    \caption{The parameter space we explore with FARGO3D. Filled circles indicate simulations which have converged to VSS ($\dot{M}$ deviations less than $10\%$). 
    Brown open squares show simulations which transitioned to an eccentric disk state, and hence will be discarded from our analysis. 
   We  omit a simulation with $q=10^{-3}$ and $\alpha=10^{-4}$ that did not converge to VSS.
    We set $h=0.05$ in all simulations. 
    We have  indicated where the thermal mass, $q=h^3$, lies \citep{2004ApJ...606..520M}, as well
    as lines of constant $K= q^2/(\alpha h^5)$. 
}
    \label{fig:param}
\end{figure}

We run a suite of FARGO3D simulations to viscous steady state.  
Figure \ref{fig:param} shows the $q$ and $\alpha$ values that we cover, with
each  filled circle representing  a converged VSS simulation.  
Table \ref{tab:sims} in Appendix \ref{sec:app_table} summarizes the simulation results.
Note that we name  simulations according to their $q$ and $\alpha$ values in an obvious
 notation; e.g., 
simulation ``q1x3a3x4'' has $q=10^{-3}$ and $\alpha=3\times 10^{-4}$.

For some purposes below, it will prove convenient to group simulations
according to their value of $K$, as simulations with similar $K$ values have similar gap depths and one-sided torques (\S \ref{sec:moderate}).
Figure \ref{fig:param} shows that our chosen parameters group into clusters
with nearly (though not identically) the same values of $K$.

The ultimate result from these simulations is 
 $\Delta T$, the values of which are listed in Table \ref{tab:sims}. They are also plotted 
 versus $K$ below (Figure \ref{fig:qscaling}).  But we refrain from a  discussion of $\Delta T$
 until after we have described the simulation results in more detail.

\subsection{Standard simulation overview} \label{sec:overview}

\begin{figure}
    \centering
     \includegraphics[trim={0.3cm 0.4cm 0.2cm 0},clip,width=0.48\textwidth]{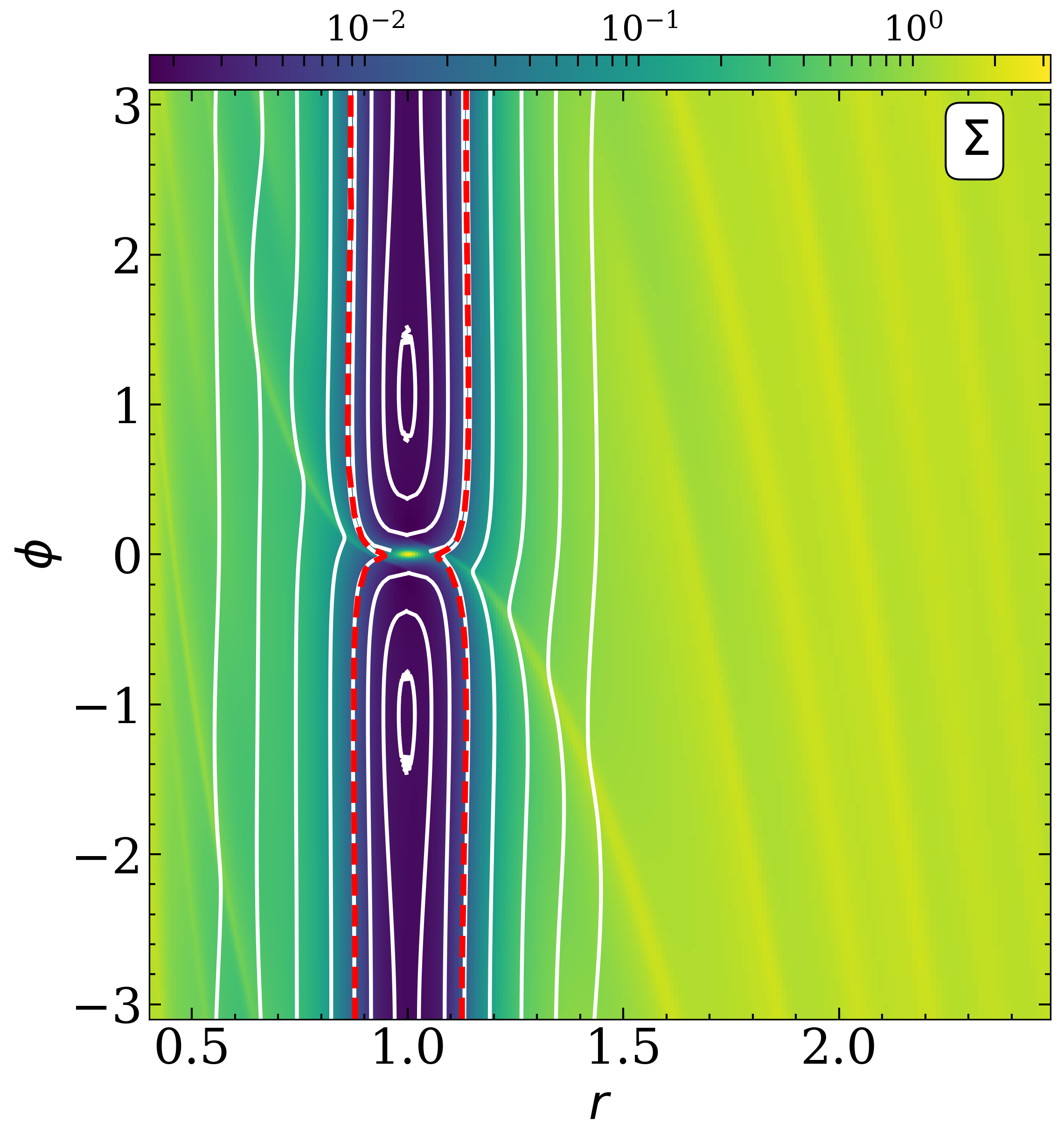}
    \caption{The two-dimensional surface density for our standard simulation with $q=\alpha=10^{-3}$. We overplot a sample of gas streamlines (white lines) and the separatrices (red dashed line) which separate the circulating streamlines from the librating streamlines. }
    \label{fig:2d}
\end{figure}

We focus first on a single ``standard'' simulation, q1x3a1x3 (i.e., $q=\alpha=10^{-3}$ implying $K=3200$). Its
pileup factor is $\Delta T/(\dot{M}\ell_p)=2.0$.
In Figure \ref{fig:2d}, we show the 2D VSS surface density for this simulation with several gas streamlines overplotted.
One may observe the deep gap ($< 0.1\%$) surrounding the planet, with  trailing spiral arms  visible in the inner and outer disks.

\begin{figure}
    \centering
    \includegraphics[trim={0.2cm 0.1cm 0.1cm 0},clip,width=0.48\textwidth]{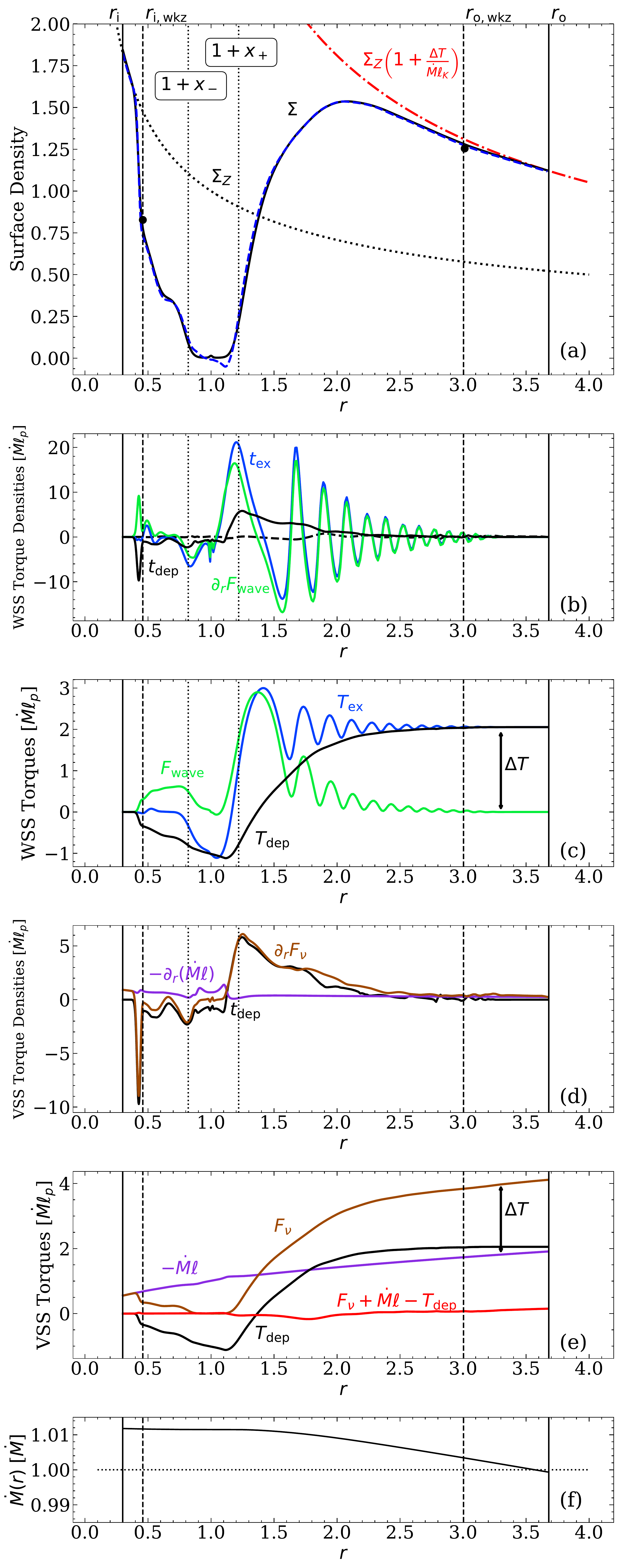}
    \caption{Summary of our standard simulation with $q=\alpha=10^{-3}$. See text for details.
    }
    \label{fig:ex}
\end{figure}

In Figure \ref{fig:ex} we show the principle torque balances from \S\ref{sec:1d} for this simulation. 
All of the quantities shown have been averaged over $3,000$ orbits of the planet.
Details of the torque calculations and averaging procedure are given in Appendix \ref{sec:app_num}. 
In each panel, the vertical solid lines mark the extent of our computational domain, and the dashed vertical lines mark the start of the wave-killing regions. 

We now walk through each of the panels. 
Panel (a) shows the azimuthally averaged $\Sigma$ profile. 
Of particular note is the gas pileup where $\Sigma$ is roughly a factor of two larger than $\Sigma_Z$. 
Panels (b) and (c) show the torques from the 
WSS equation  (Eq.\ \ref{eq:wss}).
In panel (b) we show the differential torques $t_{\rm ex}$, $t_{\rm dep}$, and $d F_{\rm wave}/ dr$ while in panel (c) we show the integrated torques (or fluxes), $T_{\rm ex} \equiv \int_0^r dr' \, t_{\rm ex}(r')$ , $T_{\rm dep} \equiv \int_0^r dr' \, t_{\rm dep}(r')$, and $F_{\rm wave}$. 
In all cases, we compute the deposited torque profile using Eq.\ \eqref{eq:wss} rather than Eq.\ \eqref{eq:tdep_def}.  
In computing the torque profiles we neglect the contribution from waves with $m=1$.\footnote{More specifically, we omit the contribution from $m=1$ to the sums defined in Appendix \ref{sec:app_num} (Eq.\ \ref{eq:fwavea}).}
We do this simply for aesthetic reasons: the $m=1$ contribution to  $t_{\rm ex}$
 is highly oscillatory, but hardly affects 
$t_{\rm dep}$
---as shown by the dashed black line in panel (b), which plots the $m=1$ contribution to $t_{\rm dep}$.

Most of the excited torque comes from near the peak of the $t_{\rm ex}$ profile. 
More precisely, the one-sided torques are dominated by the values at the inner and outer peaks of the torque per unit logarithmic distance, $x t_{\rm ex}$, where $x = r-1$ is the distance to the planet. 
For this simulation the inner and outer peaks occur at 
$x_-\approx-0.18$ and $x_+ \approx 0.23$, shown as the dotted vertical lines, respectively. 
At larger distances, $t_{\rm ex}$ becomes oscillatory due to the dominance of isolated low-$m$ Lindblad resonances. 
However, the torque from these oscillatory regions mostly cancels, as may be seen in the plot of $T_{\rm ex}$.

Panel (c) illustrates the distinction between torque excitation and deposition (\S \ref{sec:exdep}): the $T_{\rm dep}$ profile is broader than  $T_{\rm ex}$, because
waves transport angular momentum away from the planet.

\begin{figure*}
 \centering
   \includegraphics[trim={.3cm .3cm .25cm 0},clip,width=.85\textwidth]{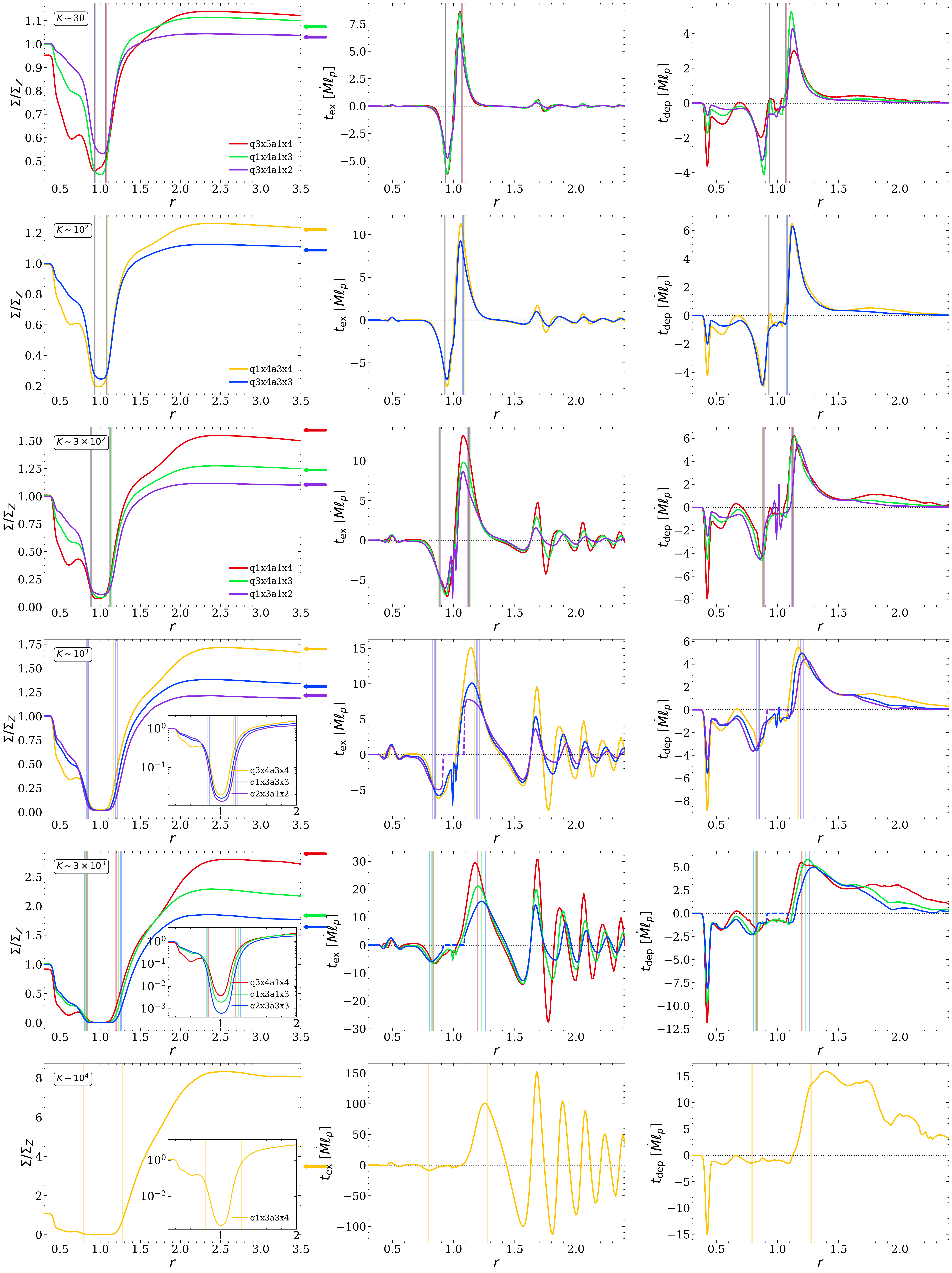}
    \caption{Surface densities (left column), $t_{\rm ex}$ profiles (middle column), and $t_{\rm dep}$ profiles (right column) for VSS simulations with $K \gtrsim 30$. 
    Each row contains simulations close to the given $K$. The line colors indicate the $\alpha$ value of the simulation. The vertical lines mark the locations of 
    the inner and outer peaks of $x t_{\rm ex}$. For $K \gtrsim 10^3$ we show an inset of the surface density profile on a logarithmic scale. The horizontal arrows to  the right  of the density plots show
    the prediction for $\Sigma/\Sigma_Z$ at $r=3.5$ that results from our fitted power-law formula for 
    $\Delta T$
      (displayed in Eq.\ \eqref{eq:dt_fit}).}
    \label{fig:kall}
\end{figure*}

Panels (d) and (e) show the torques from the VSS equation (Eq.\ \ref{eq:tdep}), with
$t_{\rm dep}$ repeated from earlier panels. That the three torques nearly sum to
zero in panel (e) illustrates that our simulation has reached VSS. In fact, the deviation from
zero in that panel is mainly due to the neglect of the $m=1$ mode. 
The detailed shape of these profiles near the planet play an important role in determining
$\Delta T$ when the gap is very deep---a point we return to in \S \ref{sec:model}.
Panel (f) shows that $\dot{M}$ is nearly constant throughout the disk, implying that
mass transport has reached steady state---in addition to angular momentum transport.

We convert the $T_{\rm dep}$ profile in panels (c) and (e)
into a $\Sigma$ profile via the VSS equation (Eq.\ \ref{eq:fnustar}), with
$F_\nu=3\pi\nu\Sigma l$, i.e., ignoring non-Keplerian contributions, and
plot the result in panel (a) as a blue-dashed line. The agreement with the 
true $\Sigma$ profile is excellent, except for a small disagreement near
the bottom of the gap where the non-Keplerian effects are evidently important.

For this simulation, the outer wave killing zone has little effect, because the waves
have already damped before reaching $r_{o,wkz}$, as evidenced from the fact
that both $t_{\rm dep}$ and $F_{\rm wave}$ are nearly zero by then. 
Conversely, the inner wave killing zone has a dramatic effect on the $\Sigma$ profile: it forces
it to rise to $\Sigma_Z$  across an artificially short distance.  But 
one may see that this artificiality has negligible effect on the value of $\Delta T$, or on quantities
such as the depth of the gap. 
 In a realistic disk with no wave killing zone and $r_i\rightarrow 0$, the
waves would deposit their angular momentum at smaller $r$, resulting in a more gradual 
rise of $\Sigma$ inwards. But the same amount of angular momentum would still be deposited, 
because our artificial wave-damping prescription conserves angular momentum.
In other words, the (non-wave-killing) computational  domain  need only capture most of the wave {\it excitation}
rather than the wave {\it deposition}, in order to correctly determine the torques, and hence $\Delta T$. To illustrate this point further, 
the black circles in panel (a) show the surface density calculated from Eqs.\ \eqref{eq:wss} and \eqref{eq:fnu1} using the values of $F_{\rm wave}$ at the wave-killing boundaries. 
These agree with the true surface density profile. 
Nonetheless, we emphasize that our $\Sigma$ profile is incorrect at $r<r_{i,wkz}$, and the 
resulting error will be seen to be more dramatic in some of our other high-$K$ simulations.

\subsection{Radial profiles at different $q$ and $\alpha$} \label{sec:profiles}

\begin{figure}
\centering
\includegraphics[trim={0.3cm 0.1cm 0.2cm 0},clip,width=0.48\textwidth]{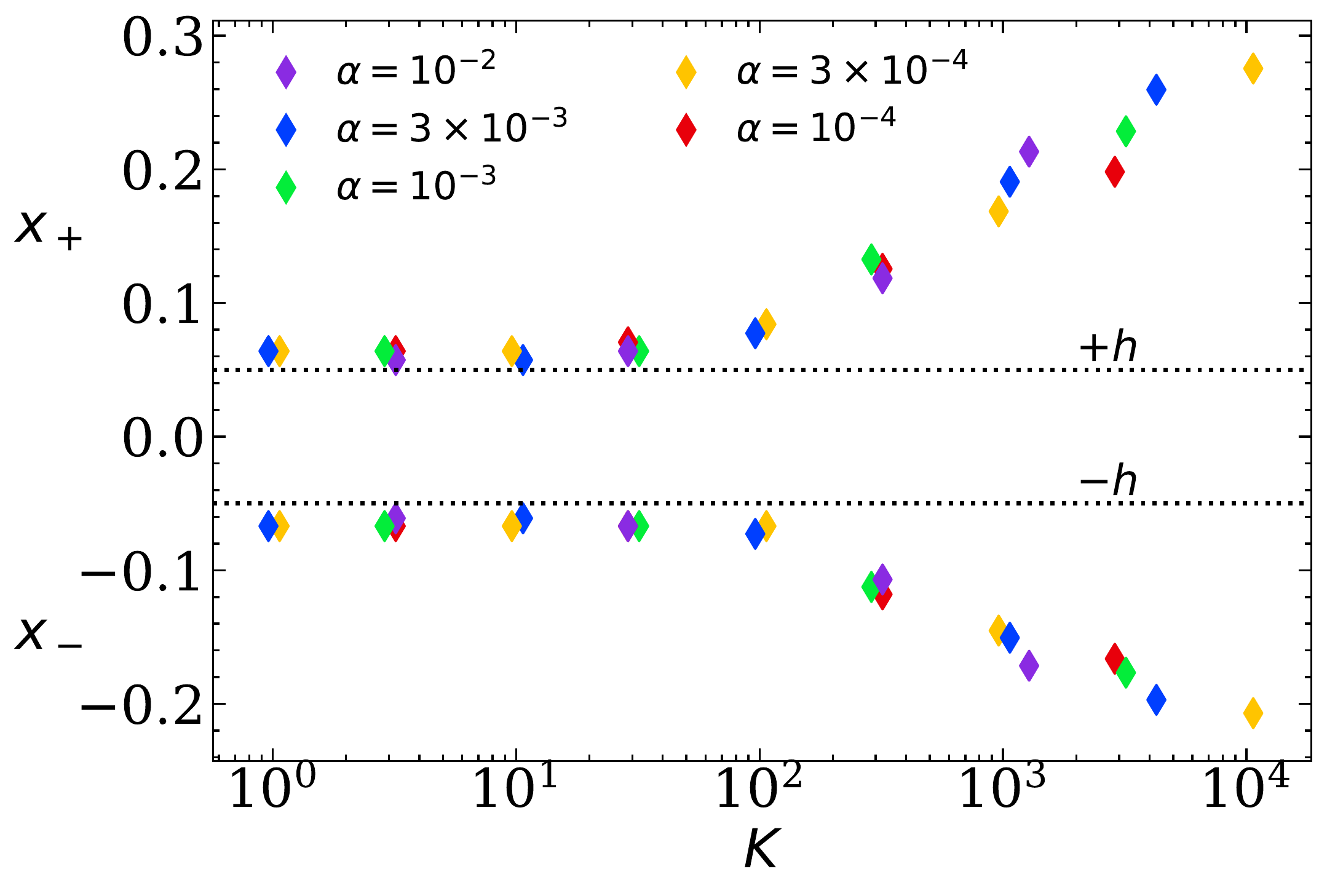}
\caption{The measured values of $x_\pm$ for all of the VSS simulations. 
The color of each point represents its $\alpha$ value.
Simulations with $K \lesssim 100$ have $|x_\pm| \approx h$ while larger $K$ simulations have $|x_\pm| > h$.
}
\label{fig:xpm}
\end{figure}

In Figure \ref{fig:kall} we show the $\Sigma/\Sigma_Z$, $t_{\rm ex}$, and $t_{\rm dep}$ profiles for our VSS simulations with $K \gtrsim 30$. 
We group simulations by their $K$, 
even though  the value of $K$ varies slightly within each group.
Each simulation is colored by its $\alpha$ value -- a scheme which we adopt for the remainder of the paper. 
In plotting the $t_{\rm ex}$ and $t_{\rm dep}$ profiles we have removed the contribution from within the planet's Hill sphere (which has a radius of $(q/3)^{1/3}$) for $q=2 \times 10^{-3}$, and replaced the missing bit 
with dashed lines. We do so  because the profiles have large spikes near the planet that 
hide
the other profiles. But these
tend to cancel and so are likely not of great importance.
One deduces the following from these figures:
\begin{itemize}
    \item $K$ is an excellent ordering parameter: simulations with similar $K$ have  similar gap depths and torque density
    profiles. This is to be expected from the theory for moderately deep gaps (\S \ref{sec:moderate}), but it continues to hold true for very deep gaps ($K\gtrsim 100$). 
    Furthermore, as expected, 
  simulations with larger $K$ tend to have both deeper gaps and larger pileups. The largest pileup we find is $\sim 10$, at $K \sim 10^4$. 
    \item As before,  vertical lines in the $\Sigma$ plots show 
the locations where inner and outer excited torques predominantly come from ($x_\pm$), 
defined as   where
  $xt_{\rm ex}$ reaches
its inner and outer extrema. We argued in \S  \ref{sec:moderate} that these
locations are of key importance.
 We see that higher $K$ systems
excite their torque farther from the planet.  Figure  \ref{fig:xpm} shows these locations
 for all of our simulations.  Evidently, simulations with
$K\lesssim 100$ have their torque excited at the torque cutoff ($h$), and hence qualify
as moderately deep gaps (\S \ref{sec:moderate}). As $K$ increases, the excitation site
is pushed further out, because the gap at $h$ becomes so deep that there is negligible
wave excitation there. 
    
    \item Simulations at a given $K$  have different $\Delta T$, as inferred from the relative
    heights of their pileups; i.e., their two-sided torque
    differs even though their one-sided torque is quite similar. This is only superficially paradoxical,   
    because the differences in one-sided torques across different simulations become amplified
    in forming the two-sided torque.  The sense of variation is that, at fixed $K$, simulations
    with lower $\alpha$  (and hence lower $q$) have larger pileups (and hence higher $\Delta T$).
   As we shall show below, this trend is systematic, and is not caused by the variation
    of $K$ within each group.      
\end{itemize}

\begin{figure*}[t]
\centering
\includegraphics[trim={0.3cm 0.1cm 0.2cm 0},clip,width=0.78\textwidth]{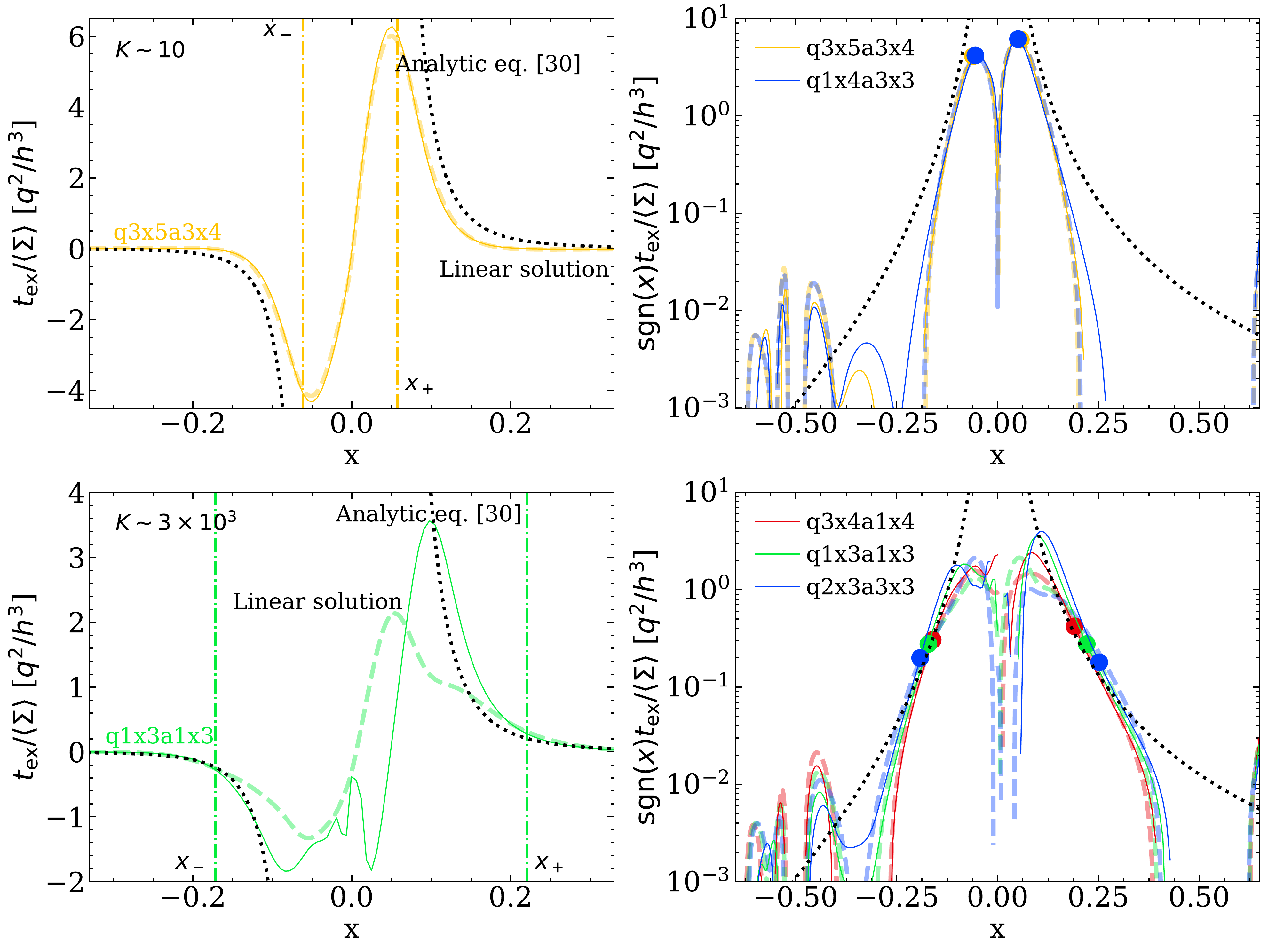}
\caption{The specific torque profiles, $t_{\rm ex}/\avg{\Sigma}$, 
for $K\sim 10$ (top row) and $K\sim 3,000$ (bottom row).
The figure shows that the linear torque is always an adequate approximation 
at $x\sim x_\pm$. 
And in the high-$K$ case, the analytic torque is a good approximation 
at $x\sim x_\pm$.
See main text for further detail.
}
\label{fig:torque_excitation}
\end{figure*}

\subsection{Torque excitation} \label{sec:excitation}

In order to calculate $\Delta T$ from first principles, one may proceed in an iterative way.  First, given 
a background surface density profile, one determines the excited torques ($T_\pm$). Second, one calculates {\it where} that excited torque is deposited, after it has been  carried further away from the planet by waves (i.e., $t_{\rm dep}$); $t_{\rm dep}$ then 
  determines the $\Sigma$ profile via the VSS equation (Eq. \ref{eq:fnu1}).
Finally, one uses that new $\Sigma$ profile to calculate the new $T_\pm$, and iterates until convergence.  
In this subsection, we focus on the first step.  In particular, we show 
 that given $\Sigma$ one may predict $T_\pm$ quite simply---without running
 a full hydrodynamical simulation.
Somewhat surprisingly, the case of a very deep gap is even simpler than that of a moderate gap.

 Following, e.g., 
 \cite{1980ApJ...241..425G} \citep[but see also][]{1993ApJ...419..155A,1993Icar..102..150K,1997Icar..126..261W,2002ApJ...565.1257T,2012ApJ...747...24R,2012ApJ...758...33P}, we 
 calculate  $T_\pm$, given $\Sigma$, under the assumption that the waves
 are linear. We therefore 
  linearize the equations of motion, 
and  solve them numerically, subject to outgoing boundary
conditions. See Appendix \ref{sec:app_linear} for details.
 This is similar to what we have done in \S \ref{sec:moderate}, except here we use the
 $\langle \Sigma\rangle$ and $\avg{\Omega}$ profiles from the hydrodynamical simulation in VSS as the background. 
The top left panel of  Figure \ref{fig:torque_excitation} compares the
 profile of 
 $t_{\rm ex}/\langle\Sigma \rangle$ from the linear solution (dashed line) with that from  
 one of our hydrodynamical simulations with $K\sim 10$. 
  The agreement is almost
 perfect, because the waves launched in the simulation are indeed linear at this modest
 value of $K$. This demonstrates that one need not solve a full hydrodynamical simulation 
 to obtain $T_\pm$ for this value of $K$---only the much simpler linear solution
 is needed (even though it is still numerical).
 The lower left panel repeats the exercise, but for a simulation with $K\sim 3,000$. 
 Now, the linear and hydro solutions disagree close to the planet, demonstrating that
 the waves are very nonlinear there.  But near where the torques are excited---i.e., 
 in the vicinity of $x_\pm$---the linear and hydro solutions agree quite well. 
 Hence for this simulation, too, the simple linear solution suffices to predict $T_\pm$, once $\avg{\Sigma}$ is specified.

 One might wish for an even simpler--- and purely analytic---prediction for
 $T_\pm$. 
 In  Appendix \ref{sec:app_ward}, we derive a simple extension to the ``standard torque
 formula'' of  \cite{1980ApJ...241..425G} that accounts for the leading asymmetry between
 inner and outer torque at large distances from the planet, starting from the more general expression 
 derived by \cite{1993ApJ...419..155A} and \cite{1997Icar..126..261W}. Our result is
\be \label{eq:tex_asymm} 
t_{\rm ex} = \pm C  \Sigma {q^2\over x^4}  \left( 1 + 2.26 x \right)  ,
\ee
where $C \approx 2.5$; the above expression is independent of $h$.
This is shown as dotted lines in the left panels of Figure \ref{fig:torque_excitation}. 
As seen in the figure, this formula fails near the torque cutoff, and hence is inadequate
to explain $T_\pm$  in the low-$K$ simulation. But it matches the high-$K$ simulation well at $x_\pm$.
To see the behavior more clearly, in the right panels of the figure we re-plot on a log-scale, 
and also
  add our other simulations at the two $K$ values.
The values of $t_{\rm ex}/\avg{\Sigma}$ at $x_\pm$ (indicated by the circles) are close to the values from Eq.\ \eqref{eq:tex_asymm} near $x_\pm$. 
Therefore, for high-$K$ one may predict $T_\pm$  by multiplying  Eq.\ (\ref{eq:tex_asymm}) at $x_\pm$ by the surface density at those
locations ($\Sigma_\pm$):
\be \label{eq:tpm_model}
T_\pm \approx \pm C \Sigma_\pm \frac{q^2}{|x_\pm|^3} \left(1 + 2.26 x_\pm \right) \ ,
\ee
after dropping an order-unity coefficient. 
 In \S\ref{sec:model} below, we compare this prediction for $T_\pm$ with the actual values 
   in all of our $K\gtrsim 100$ simulations.

To summarize, we have shown that the linear calculation suffices to determine $T_\pm$ in 
all of our simulations, and the much simpler
 standard torque formula (with the added asymmetry) suffices for the high-$K$ simulations. 
The latter result might appear surprising in light of studies showing that the standard
torque formula can be quite inaccurate---it can even give the wrong sign for  $t_{\rm ex}$ at
certain distances from the planet \citep[e.g.,][see also the right panels of Figure \ref{fig:torque_excitation} at $|x|\gg |x_\pm|$]{2011ApJ...741...56D,2012ApJ...747...24R,2012ApJ...758...33P}. Nonetheless, those inaccuracies evidently do not 
have a large effect on $T_\pm$---at least for the range of parameters spanned by our simulations.

\begin{figure*}
\centering
\includegraphics[trim={.3cm .3cm .2cm 0},clip,width=.95\textwidth]{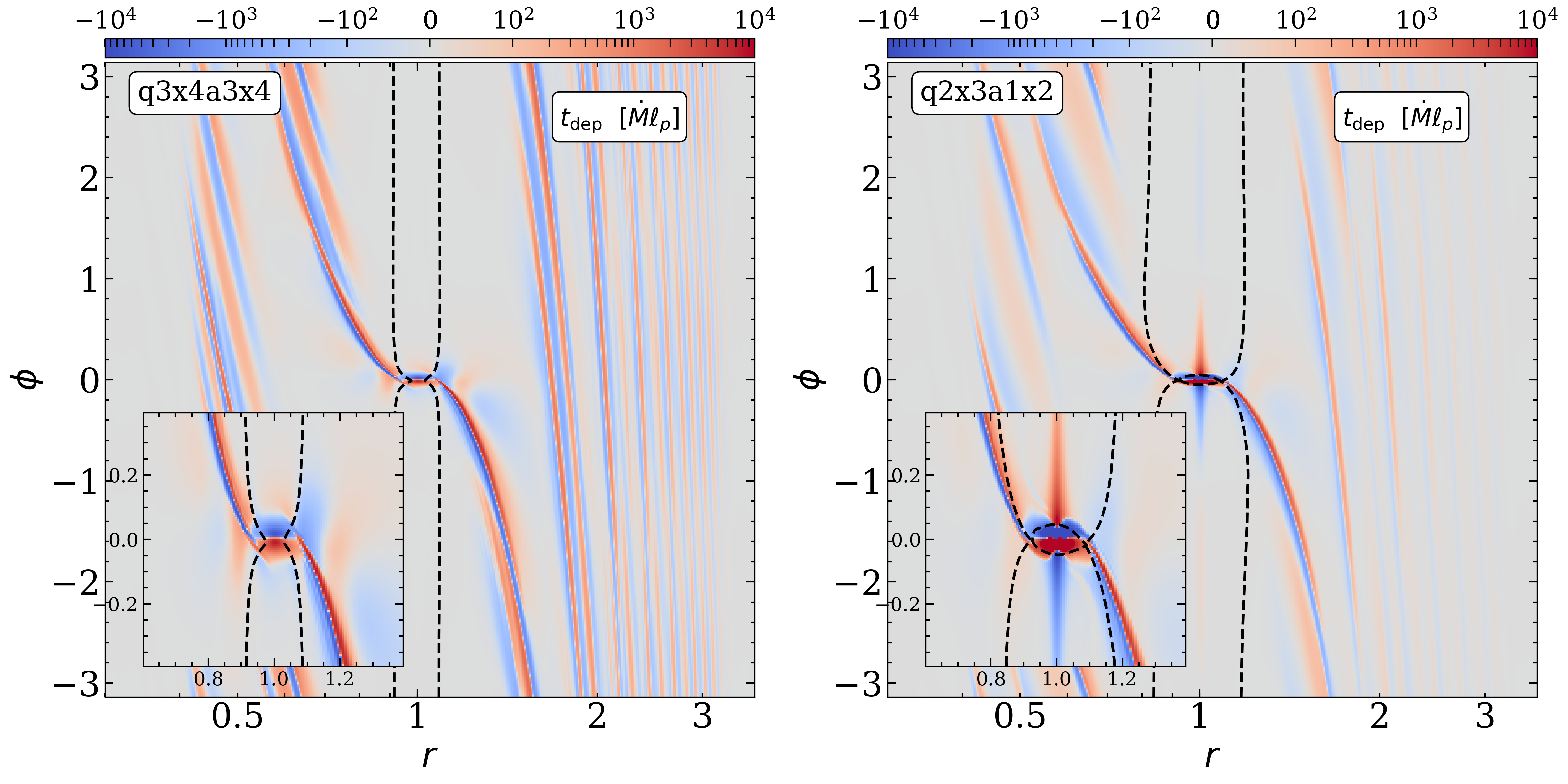}
\caption{Two-dimensional maps of $t_{\rm dep}$ for two simulations with $K \sim 1,000$. The colorscale is logarithmic for values greater than $100$ and linear for values less than $100$. The separatrices (black lines) mark the transition from librating to circulating fluid streamlines in the co-rotating frame. The separatrix centered on the planet marks the extent of the circumplanetary disk region where fluid elements orbit the planet.}
\label{fig:tdep2d}
\end{figure*}

\subsection{Separating Lindblad from Co-orbital Torques}

 We shall separate Lindblad from co-orbital torques in the simulations, 
in order to show  that (i) Lindblad torques are well-understood  for
moderate gaps, and (ii)  co-orbital torques are usually sub-dominant, across all simulations.
 Before doing so, we describe here how we separate out the two torques.

Previous treatments have separated the torques by calling 
 torque excited inside the horseshoe zone the co-orbital torque, and  that excited outside the
 Lindblad torque \citep[e.g.,][]{2009MNRAS.394.2283P}.  However, by examining 2D plots of $t_{\rm ex}$ (not shown), such
a distinction appears ambiguous: there is no clear boundary separating one type of torque from
another.  Instead, we have found that the distinction becomes much clearer when examining
 $t_{\rm dep}$.
 Figure \ref{fig:tdep2d} shows 2D maps of $t_{\rm dep}$  for two simulations with $K \sim 1,000$. 
The black dashed lines show the separatrices.
These mark the transition from librating to circulating streamlines in the planet's co-rotating frame. 
The zoomed-in insets  of Figure \ref{fig:tdep2d} show that the distinction between
 Lindblad and co-orbital torques is quite apparent: Lindblad torques 
 show up as outwardly projecting arms, and co-orbital torques as the nearly elliptical structure
 near the planet (caused by the U-turn of fluid that follows nearly horseshoe orbits).
 One may understand  why $t_{\rm dep}$ is more useful for separating out the
 two torques as follows: before contributing to $t_{\rm dep}$, the Lindblad torque propagates away
 from its point of excitation along the spiral arms, away from the co-orbital zone.
 Figure \ref{fig:tdep2d} also shows that the separatrix is only an approximate dividing
 line between the spiral-type and elliptical-type pattern.
  As such, we separate the two  contributions by eye for each simulation.

\subsection{Total torques} \label{sec:torques}

\begin{figure*}
\centering
  \includegraphics[trim={.3cm .3cm .2cm 0},clip,width=.95\textwidth]{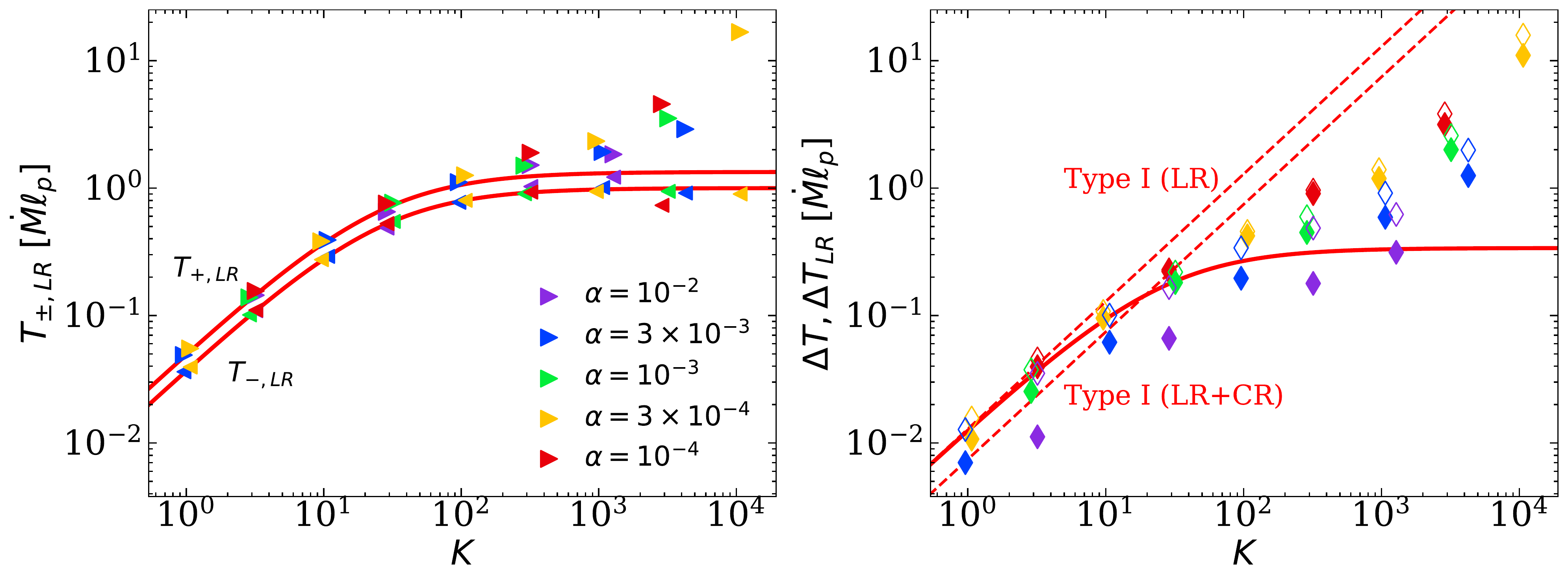}
\caption{One-sided Lindblad torques, $T_{\pm,LR}$, (left) and  total (two-sided) torques, $\Delta T$, (right) for all of our VSS simulations as a function of $K$. 
In the right panel, the open points show the total Lindblad torque and the filled points show the total torque (Lindblad $+$ co-orbital); the latter is what is relevant for the pileup and planet migration.
The solid red lines in both panels show the predicted Lindblad torques
for moderately deep gaps from Eqs.\ \eqref{eq:Tpm} - \eqref{eq:dt1}, showing excellent
agreement with the Lindblad torques (open diamonds in the right plot) from the simulations at $K\lesssim 100$.
The dashed red lines in the right panel show the linear torque scaling from \citet{2002ApJ...565.1257T}. 
The upper line is the purely Lindblad torque while the bottom line is the Lindblad $+$ co-orbital torque. 
}
\label{fig:T2s}
\end{figure*}

\begin{figure}
\centering
\includegraphics[trim={.3cm .3cm .2cm 0},clip,width=.48\textwidth]{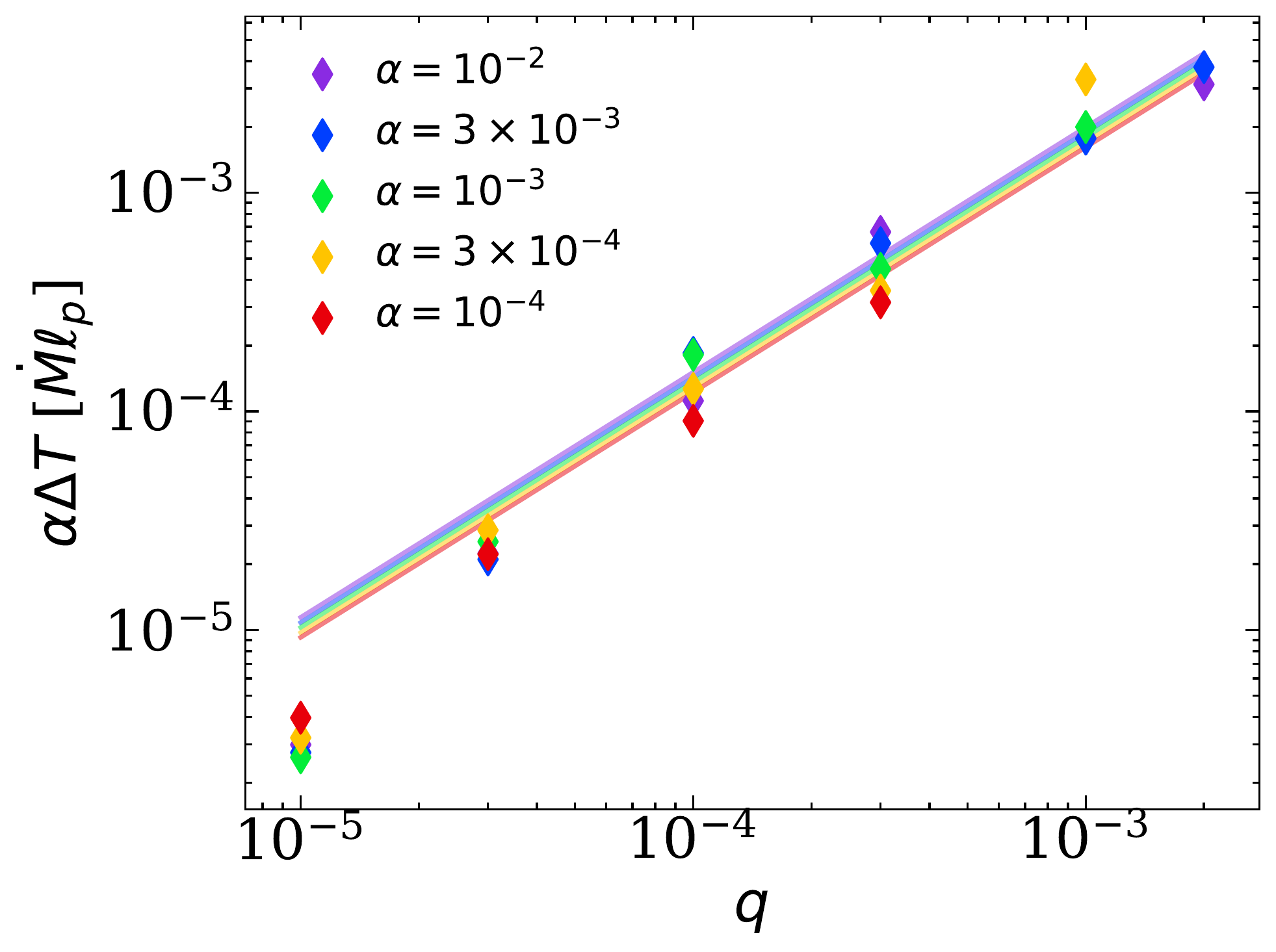}
\caption{$\alpha \Delta T$, normalized to $\dot{M} \ell_p$, as a function of $q$ for all of our VSS simulations. 
The lines show the result of the fit given in Eq.\ \eqref{eq:dt_fit} for each of the $\alpha$ values, demonstrating that $\alpha \Delta T$ is roughly independent of $\alpha$.
}
\label{fig:qscaling}
\end{figure}

We  present here the torques from our suite of VSS simulations.   
Figure \ref{fig:T2s} shows the measured torques as a function of $K$. 
The left panel shows the one-sided inner (leftwards pointing triangles) and outer Lindblad torques (rightwards pointing triangles). 
The right panel  shows the final $\Delta T$ (solid points),
which is what is relevant
for the pileup and the planet's migration;
 it also 
shows the Lindblad component of $\Delta T$ (open points).  
For the most part, the co-orbital torques constitute a tens of percent correction to the total $\Delta T$. 
However, they are significantly more   important 
in the high $\alpha$ simulations ($\alpha\gtrsim 3\times 10^{-3}$).

In both panels, we plot as solid red lines our ``moderate gap'' predictions for the Lindblad torques
from \S\ref{sec:dt_theory}. The agreement is excellent at $K\lesssim 100$, both for the
one-sided torques, and for the Lindblad component of $\Delta T$. 
We also show in the right panel  the canonical Type I scaling (which corresponds to the no gap limit, $K\ll 100$) as the dashed red lines, taken from \cite{2002ApJ...565.1257T}. The upper one
is for Lindblad only, and agrees with the prediction (and simulations).  The lower one includes
co-orbital torques, and appears to provide 
a better match to the low $K$ simulations\footnote{
Our $K\ll 100$ Lindblad prediction for $\Delta T$ is the same as that of \cite{2002ApJ...565.1257T} 
because the $C_\pm$ we calculated from linear theory (Eqs.\ \ref{eq:cp}--\ref{eq:cm}) differ negligibly from theirs. That same linear calculation also produces the corotation torque, and we have verified that we get the same
result as  \cite{2002ApJ...565.1257T} for that as well.
}.

Proceeding to  $K\gtrsim 100$, we see that the one-sided  inner torques asymptote to values of $T_{-,LR} \approx -\dot{M} \ell_p$, while $T_{+,LR} > \dot{M} \ell_p$. 
In VSS, we know that the inner torque cannot exceed $\dot{M} \ell_p$ and deviates from $\dot{M} \ell_p$ by the gap depth (Eq.\ \ref{eq:fnu1} at $r=1$), which for our highest
$K$ simulations is less than $0.1\%$ (Figure \ref{fig:kall}).
Unlike the inner torque, the outer torque has no such restriction.
Clearly, the moderate-gap $\Delta T$ prediction is no longer valid once $K \gtrsim 100$, as $\Delta T$ continues to increase with $K$ sub-linearly. 
By contrast,  \citet{2018ApJ...861..140K} find that $\Delta T$ is nearly constant (for fixed $h$) at those $K$ values. 
As we show below in \S\ref{sec:other_dts}, this discrepancy is due to the \citet{2018ApJ...861..140K} results not being in VSS.

 From the right panel,
 we see that the $\Delta T$'s from simulations with the same $\alpha$ (i.e., 
with the same color) trace out distinct lines, with the height of the line
dropping with increasing $\alpha$. This implies that 
there is variation  in  $\Delta T$ at fixed $K$, with higher $\alpha$ simulations  having lower $\Delta T$, as already  suggested by  the pileups
seen in 
Figure \ref{fig:kall}. 
Motivated by this, 
we fit $\Delta T$ to a power law in $q$ and $\alpha$ above $q=10^{-4}$. 
The result is,
\be \label{eq:dt_fit}
 \Delta T = 4.3_{2.8}^{6.6} q^{1.05 \pm 0.06} \alpha^{-0.91 \pm 0.04}  \dot{M} \ell_p ,
\ee
 where the errors are statistical.
We have chosen to omit the $K \sim10^{4}$ point from our fit as it has an unusually large $\Delta T/(\dot{M} \ell_p)$. 
Given that the dependence is nearly $\propto \dot{M}\ell_p\alpha^{-1}$, 
we show in Figure \ref{fig:qscaling}, $\Delta T\times\alpha$ as a function of $q$ (not $K$). 
The lines in the figure show the best fit $\Delta T$ values for each of the $\alpha$ values.

Given $\Delta T$, we can calculate the expected pileup magnitude and $\Sigma$ profile outside of the location where $T_{\rm dep} \approx \Delta T$ using Eq.\ \eqref{eq:sigma_sol}, 
\be \label{eq:pileup}
\Sigma \approx \Sigma_Z \left(1 + \frac{\Delta T}{\dot{M} \ell} \right) .
\ee
In the surface density profiles of Figure \ref{fig:kall} we show the value of the pileup at r=3.5 calculated from our $\Delta T$ scaling (Eq.\ \ref{eq:dt_fit}) with horizontal arrows. 
These are in good agreement with the true $\Sigma$ values from the simulations.

\section{Additional Features of the VSS Solutions} \label{sec:discussion}

\subsection{Planet Migration Rate and Validity of the VSS Assumption} \label{sec:mig}

The two-sided torque $\Delta T$ must come at the expense of the
planet's angular momentum. Hence
 for positive $\Delta T$, the planet migrates inwards at a rate
\be
{\dot{r}_p\over r_p}= - 2 { \Delta T\over M_p \ell_p} . \label{eq:mrg}
\ee
From the results of \S\ref{sec:results},  there are two different regimes for $\Delta T$: a moderate gap regime, and a deep gap regime.

For moderate gaps,  $\Delta T$ is given by Eq.\  (\ref{eq:dt1})\footnote{We again ignore the contribution from co-orbital torques as they are a minor correction (Figure \ref{fig:T2s}).},
 which  we rewrite as 
 \be
 \Delta T &=& {C_++C_-\over 1+{|C_-|\over 3\pi}K}{q^2\over h^3}\ell_p^2\Sigma_{Z,p} \\
 &=&  {2.4\over 1+.04K}{q^2\over h^2}\ell_p^2\Sigma_{Z,p} \ ,
 \ee
 where the latter expression is specialized to $h=0.05$. 
The migration rate is therefore
\be
 \frac{\dot{r}_p}{r_p} = - \frac{2.4/h^2}{1 + 0.04 K}\frac{M_d M_p}{M_\star^2} \frac{1}{\tau_{\rm orb}} \ 
 {\rm (moderate\  gaps)} ,
\ee
where $M_d = 4 \pi r_p^2 \Sigma_{Z,p}$ is a measure of the local disk mass and $\tau_{\rm orb} = 2 \pi /\Omega_p$ is the orbital period of the planet.
This is very similar to the standard Type I rate, aside from the extra gap reduction
factor in the denominator. More precisely, the standard Type I rate has
a coefficient of $2.3$  at $h=0.05$ if one ignores co-orbital torques; co-orbital torques change it to 1.6.
 \citep[e.g.,][]{2002ApJ...565.1257T,2012ARA&A..50..211K}.
Our new Type I migration rate includes the reduction effect of the gap; from Figure \ref{fig:T2s}, it is valid at $K \lesssim 100$. 

In the deep gap regime the situation is quite different, as the torque scaling switches to Eq.\ \eqref{eq:dt_fit} for the $K\gtrsim 100$ simulations that we have run (up to $K\lesssim 10^4$).
Using that fit to $\Delta T$,  the migration rate is
\be \label{eq:mig_rate}
    \frac{\dot{r}_p}{r_p} = -0.1 \frac{M_d}{M_\star} \frac{ q^{0.05} \alpha^{0.09} }{\tau_{\rm orb}} 
 \      {\rm (deep\  gaps)} ,
\ee
again for $h = 0.05$. 
Remarkably, we find that in VSS planets migrate at a rate which is roughly independent of their mass and the disk's viscosity and is only dependent on the disk-to-star mass ratio.

It is instructive to compare the VSS migration rate above with 
prior Type II results.
These typically predict that the planet migrates at the same rate as the disk's viscous accretion rate,
although sometimes with an additional  mass reduction factor when the disk is less massive than the planet \citep{1995MNRAS.277..758S,1997Icar..126..261W,1999MNRAS.307...79I,2007ApJ...663.1325E,2010apf..book.....A}.   
The general migration rate expression (Eq.\ \ref{eq:mrg}) may be rewritten as
\be
{\dot{r}_p\over r_p}= - {1\over \tau_{\rm visc}}{M_d\over M_p}{3\over 2}
\left({ \Delta T\over \dot{M} \ell_p}\right) , \label{eq:tii} 
\ee
where the disk's viscous time is $\tau_{\rm visc}=r_p^2/\nu_p$ and the bracketed factor is the dimensionless pileup factor determined from our simulations (Figure \ref{fig:T2s}).  
Therefore the planet's migration rate differs from the disk's viscous accretion rate by two factors:
$M_d/M_p$ and the pileup factor.

We conclude this subsection by examining the criterion for VSS to be valid.  The basic
assumption for VSS is that the planet migrates more slowly than the disk material
\citep[see also][]{2012MNRAS.427.2660K,2012MNRAS.427.2680K}. 
For pileup factors ($\Delta T/(\dot{M}\ell_p)$) that are of order a few or less, one therefore requires
$|\dot{r}_p/r_p|\lesssim 1/\tau_{\rm visc}$, which implies from Eq.\ (\ref{eq:tii}) that
$M_d\lesssim M_p \left( \Delta T/\dot{M}\ell_p  \right)^{-1}$. In other words, for order-unity
pileups the VSS assumption is valid when the disk is 
less massive than the planet. 
For much larger pileups there is a more stringent constraint, because material at the peak of
the pileup moves more slowly that $r_p/\tau_{\rm visc}$. But since the biggest pileups that we
have found are $\sim 10$, we shall not consider very large pileups here.

\subsection{Gap depth and width} \label{sec:gap}

\begin{figure}
    \centering
    \includegraphics[trim={0.25cm 0.2cm 0.15cm 0},clip,width=0.48\textwidth]{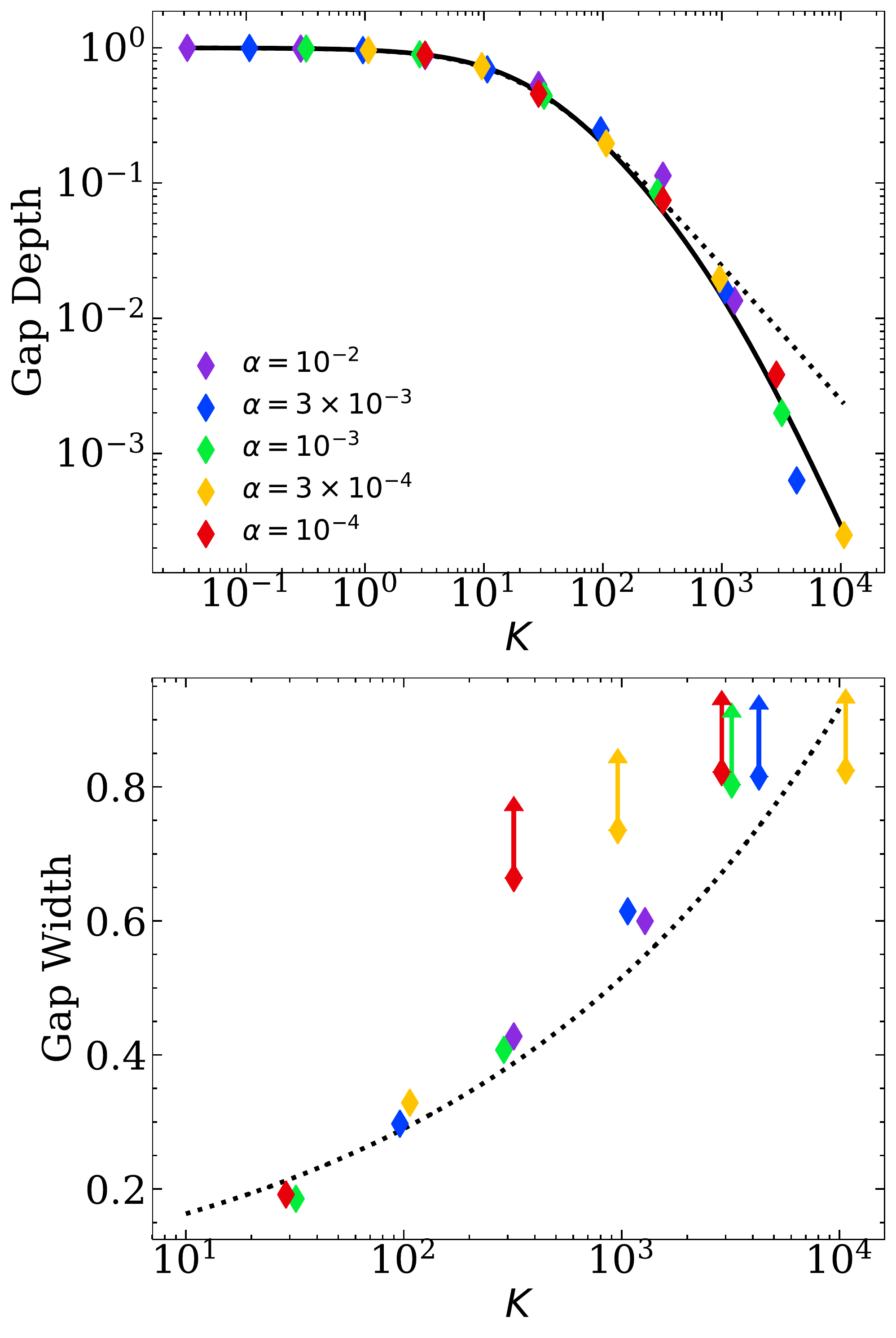}
    \caption{Top: Gap depths for all of our VSS simulations.
    Here we define the gap depth as the minimum surface density excluding the circumplanetary disk region. 
    Overplotted we show the literature scaling relation for moderately deep gaps, $1/(1 + 0.04 K)$ (dotted line; Eq.\ \ref{eq:sigsig}), and a fit to our results following the function $1/(1 + 0.04  K+ (K/K_c)^2)$, with $K_c = 180$ (solid line).
     Bottom: The gap widths for our simulations. The gap width is defined as the extent of the $\Sigma$ profile where $\Sigma < 0.5 \Sigma_Z$. 
    Arrows indicate simulations where the inner gap extends past the inner wave-killing boundary, and thus these points only correspond to lower limits.
     The dotted line shows the empirical scaling relation found by \citet{2016PASJ...68...43K}. }
    \label{fig:gap_scaling}
\end{figure}

In this subsection we compare our gap depths and widths to  previously published results, 
  and provide a new gap depth scaling that matches our VSS simulations. 

The top panel of Figure \ref{fig:gap_scaling} shows the gap depths for all of our simulations.
To calculate the gap depth we first remove the circumplanetary disk region, which  we crudely define as the region where both $|x|$ and $|\phi|$ are $< {\rm max}(h,(q/3)^{1/3})$, and then take the azimuthally-averaged surface density at the planet.
The dotted line shows the prediction for moderately deep gaps, $\Sigma_p = 1/(1 + 0.04 K)$ from Eq.\ \eqref{eq:sm1}, 
which agrees with previous studies \citep[e.g.,][]{2015MNRAS.448..994K,2015ApJ...807L..11D,2017PASJ...69...97K}. 
Above $K \sim 300$ our gap depths are significantly deeper than 
Eq.\ \eqref{eq:sm1} due to the separation of $|x_-|$ and $h$ (cf. Figure \ref{fig:xpm} and  \S\ref{sec:model}). 
A similar ``two-step" effect has also been seen for gaps around low-mass ($q \lesssim 10^{-5}$),  low-viscosity ($\alpha \lesssim 10^{-3}$) planets \citep{2018MNRAS.479.1986G}.
We fit our gap depths to a corrected scaling relation, $\Sigma_p = 1/(1 + 0.04 K + (K/K_c)^2)$, where $K_c=180$ is a fit parameter.

The bottom panel of Figure \ref{fig:gap_scaling} shows gap widths, defined as the width of the region where
 $\Sigma_p < 0.5 \Sigma_{Z,p}$.
\citet{2016PASJ...68...43K} \citep[see also][]{2017PASJ...69...97K} empirically determined that the gap width follows $\Delta_g = 0.41 (h^2 K)^{1/4}$, which we show as the dotted line. 
For low viscosity disks, we find more radially extended gaps than 
predicted from that relation. 
From Figure \ref{fig:kall},  we see that such ``extra wide'' gaps are very asymmetric with respect to the planet.
In fact, for many of our high-$K$ simulations we find that the inner boundary of the gap extends past our inner wave-killing boundary, and hence in reality could be
much wider than found in our simulation (as explained at the end of  \S \ref{sec:overview}). 

An alternative empirical gap depth and width relation that separates the $q$ and $\alpha$ dependence has recently been developed by \citet{2019arXiv190611256D}. 
We find that his relation matches our depths in the low and intermediate $K$ regime, and, in particular, reproduces the variation seen at fixed $K$.
Only at our highest $K$ and lowest $\alpha$ values ($\alpha=10^{-4}$ at $K \sim 3000$ and $\alpha = 3 \times10^{-4}$ at $K \sim 10^4$), does his relation overpredict the depth of the gap.

\subsection{Comparison with $\Delta T$'s from previous work} \label{sec:other_dts}
\begin{figure}
    \centering
    \includegraphics[trim={0.3cm 0.1cm 0.2cm 0},clip,width=0.48\textwidth]{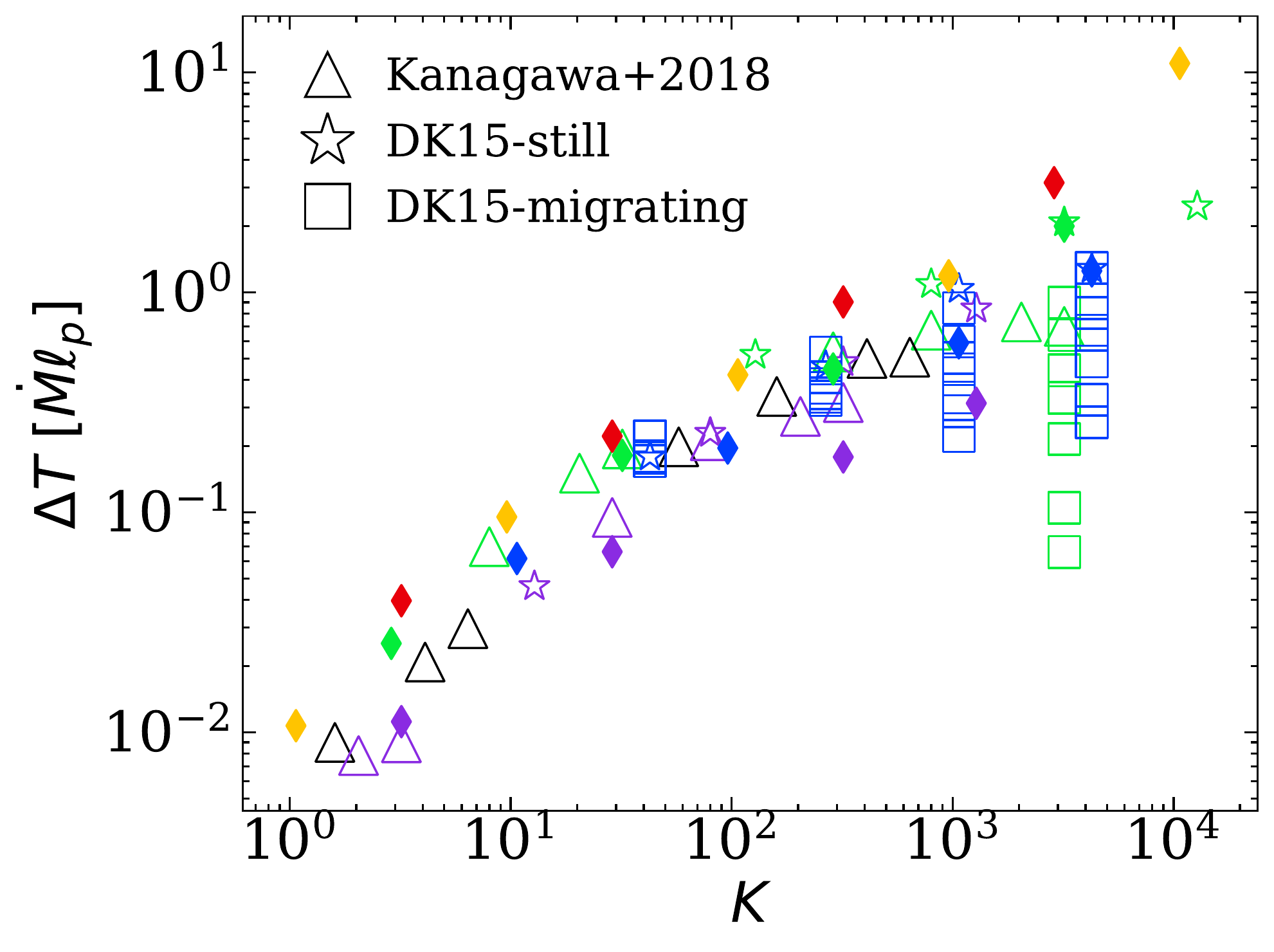}
    \caption{Comparison of our $\Delta T$ values to 
     those from \citet{2015A&A...574A..52D} (stars) and 
    the $h=0.05$ simulations from \citet{2018ApJ...861..140K} (triangles). 
     The  \citet{2018ApJ...861..140K} results, and some of those from 
    \citet{2015A&A...574A..52D}, are for migrating planets,
    while our results  
    are for stationary planets. }
    \label{fig:comp}
\end{figure}

In \S \ref{sec:1d}, we argued that simulations must have correct boundary conditions in order
to produce the correct pileup in surface density.  But we have also argued (\S \ref{sec:results} and \S\ref{sec:model}) that the value of 
$\Delta T$ should be set largely by what happens where the torque is excited ($x_\pm$), which
occurs quite close to the planet.  Therefore, prior simulations that did not adopt correct boundary
conditions might produce values of $\Delta T$ that are comparable to ours. 

Figure  \ref{fig:comp} compares our values of $\Delta T$, with those from simulations by
 \citet{2015A&A...574A..52D} and \citet{2018ApJ...861..140K}. 
In contrast to our VSS boundary conditions, \citet{2018ApJ...861..140K} set all fluid quantities equal to their initial conditions, which corresponds to the ZAM solution, at the outer boundary, and use an open boundary condition at the inner boundary. 
They also use wave-killing regions near both boundaries where they damp all quantities to their initial conditions. 
\citet{2015A&A...574A..52D} also fix all quantities to their initial conditions at the outer boundary, but fix only $v_r$ and $v_\phi$ to their initial conditions at the inner boundary. 
For their wave-killing prescription, they damp only $v_r$ and $v_\phi$ to their initial conditions and not $\Sigma$.
An additional difference is that  \citet{2018ApJ...861..140K} allow their planets to migrate, as
do  \citet{2015A&A...574A..52D} for a subset of their simulations.

Focusing on $K$ values above $100$, we see that  the non-migrating \citet{2015A&A...574A..52D}  torques are quite  close to our VSS torques with the exception of the $K \sim 1,000$ planets, suggesting
that the boundary conditions are not essential to achieving the correct $\Delta T$,  for these $K$ values.
By contrast, the \citet{2018ApJ...861..140K} torques tend to be quite different than our VSS torques, 
as do the migrating planets of  \citet{2015A&A...574A..52D}, suggesting that those migrating planets
are not in VSS---likely due to the disk-to-planet mass ratio being too large
 (\S\ref{sec:mig}). 

\begin{figure}
    \centering
    \includegraphics[trim={0.3cm 0.1cm 0.2cm 0},clip,width=0.48\textwidth]{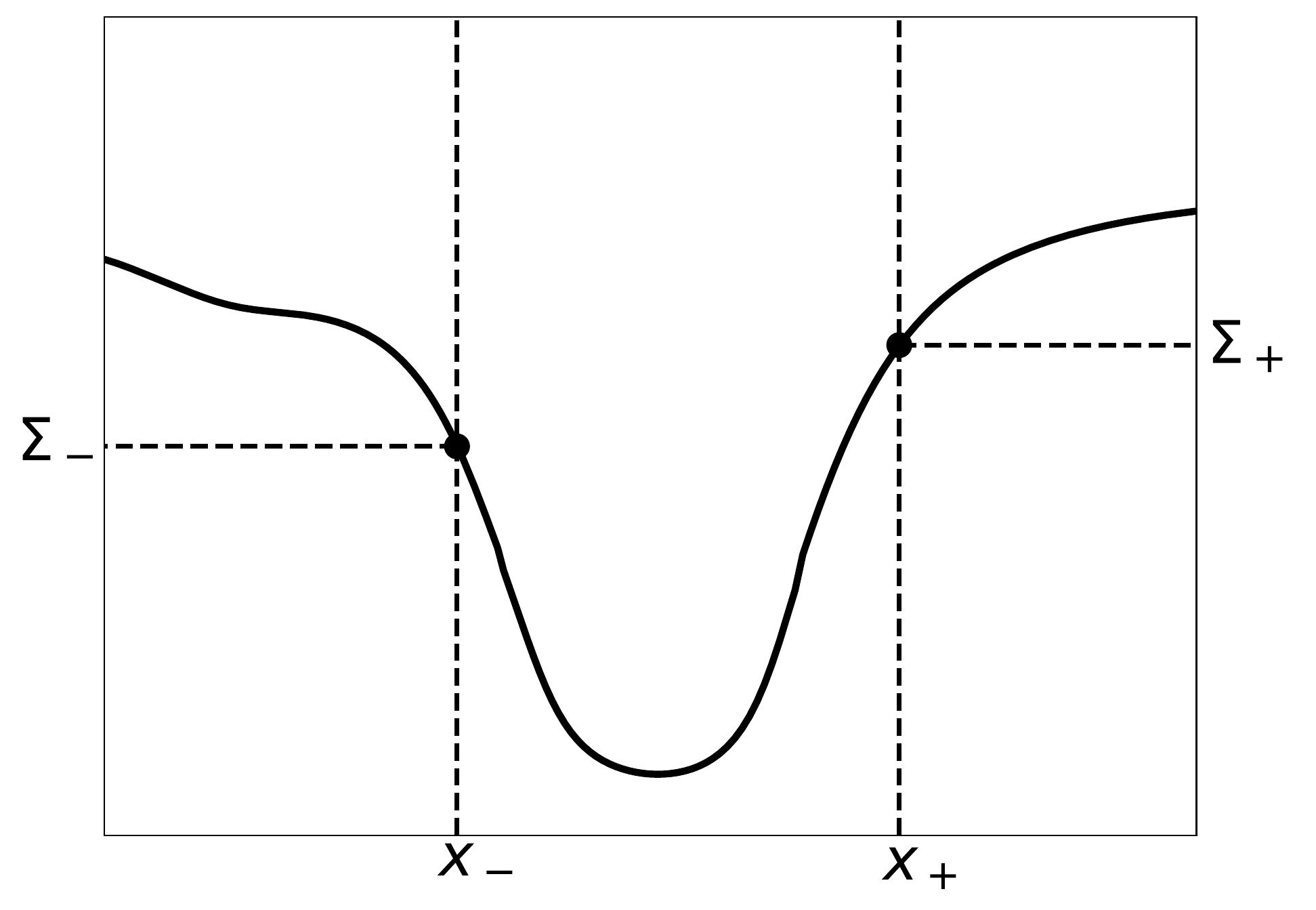}
    \caption{
    Schematic illustration of a VSS $\Sigma$ profile highlighting the four important quantities, $x_\pm$ and $\Sigma_\pm$. 
    Most of the one-sided torques are excited at $x_\pm$ with strengths given by Eq.\ \eqref{eq:tpm_model}.
    The $\Sigma$ profile between $x_-$ and $x_+$, as well as the locations of $x_\pm$ are set by the local torque deposition profile in the gap. }
    \label{fig:pm_schematic}
\end{figure}

\begin{figure}
\centering
   \includegraphics[trim={.3cm .3cm .2cm 0},clip,width=.45\textwidth]{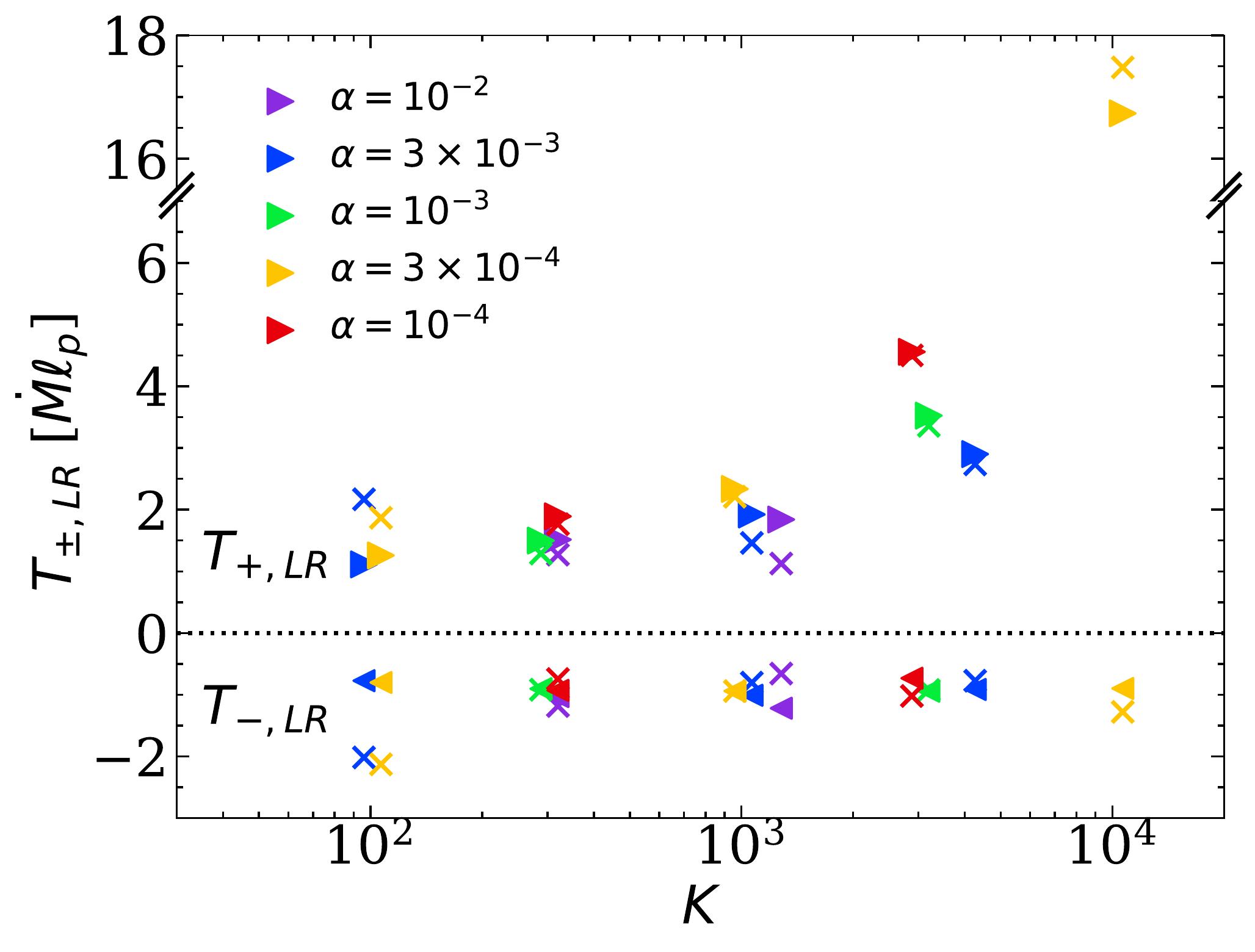}
\caption{One-sided Lindblad torques for simulations with $K \gtrsim 100$ (triangles) compared to Eq.\ \eqref{eq:tpm_model} for the measured values of $x_\pm$ and $\Sigma_\pm$ (crosses). 
Above $K \sim 100$, the agreement between Eq.\ \eqref{eq:tpm_model} and $T_{\pm,LR}$ shows that the analytic torque formula evaluated at $x_\pm$ is a good approximation to the one-sided Lindblad torques in deep gaps. 
}
\label{fig:tpm_model}
\end{figure}

\begin{figure}
    \centering
    \includegraphics[trim={0.2cm 0.1cm 0.2cm 0},clip,width=0.48\textwidth]{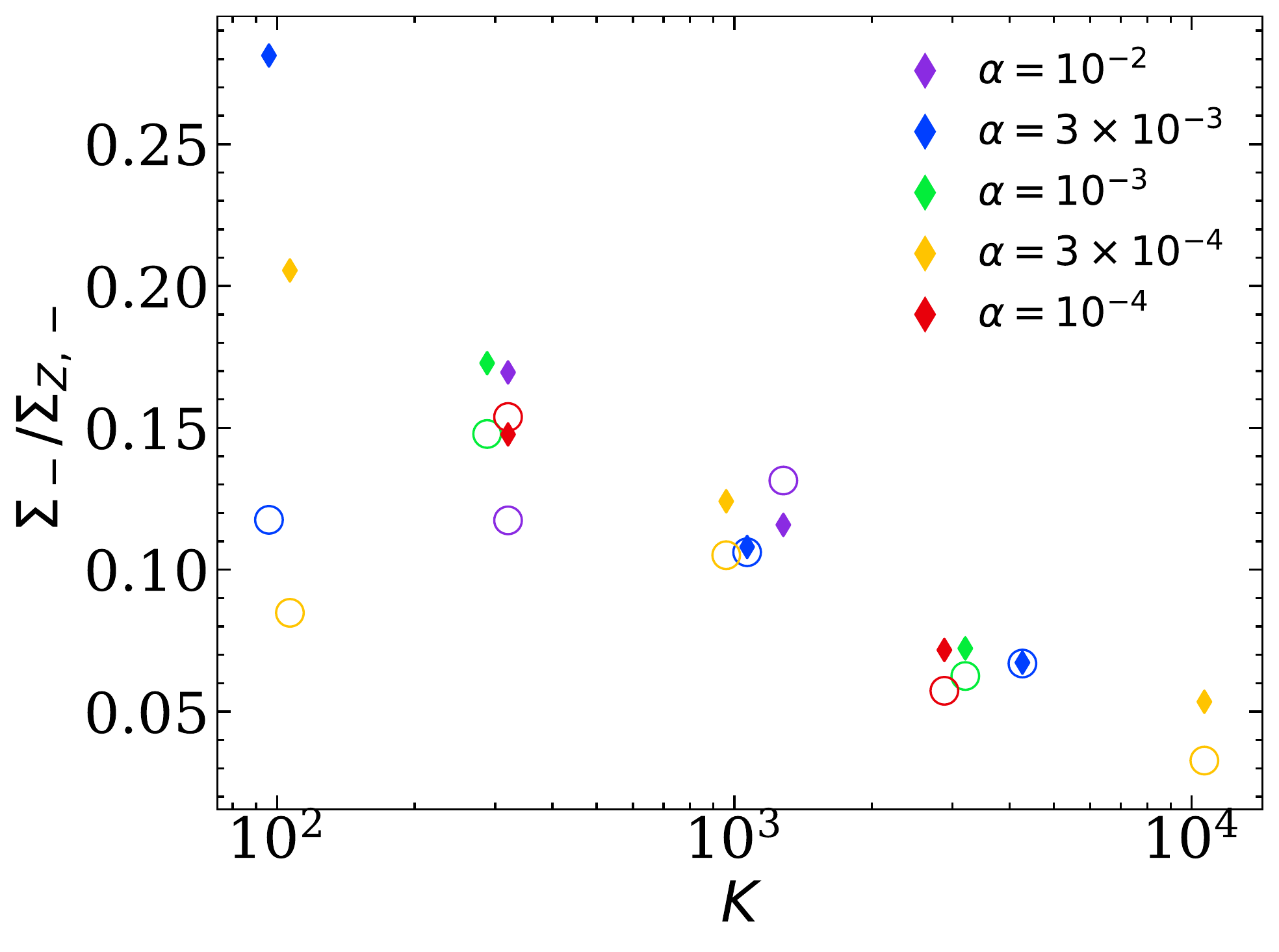}
    \caption{The $K \gtrsim 100$, $\Sigma_-/\Sigma_{Z,-}$ values (filled diamonds) compared to the values given by Eq.\ \eqref{eq:sm_high} (open circles) for the measured value of $x_-$. The agreement between the filled diamonds and open circles again lends evidence to our claim that the majority of the one-sided (inner torque) is excited at $x_-$ following the analytic specific torque density of Eq. \eqref{eq:tex_asymm}.}
    \label{fig:sm_model}
\end{figure}

\subsection{Towards a Theory of Very Deep Gaps} \label{sec:model}

The theory for moderately deep gaps  (\S \ref{sec:dt_theory})
assumes that torque excitation happens at the torque cutoff, i.e.,  that $|x_\pm| \sim h$. 
But once the gap becomes sufficiently deep, 
 torque is mostly excited further away from the planet, 
 i.e., in the gap wall, 
 where the surface density
 is higher. 
As mentioned previously, to predict where the gap wall occurs requires knowing where
torque is {\it deposited}, because that determines $\Sigma$ via the VSS equation
(Eq.\ \ref{eq:fnu1}). 
But torque deposition is difficult to calculate from first principles.  Therefore we cannot yet
 present a complete theory for very deep gaps. 
Nonetheless, we are able to take a few steps towards such a theory. 

First, as suggested in \S \ref{sec:excitation}, one may obtain the excited torques quite simply given the $\Sigma$ profile. 
In fact, one only needs four key numbers characterizing the $\Sigma$ profile: the aforementioned $x_-$ and $x_+$, as well as the $\Sigma$ values at $x_\pm$, i.e.,  $\Sigma_-$ and $\Sigma_+$.
We illustrate these important quantities in Figure \ref{fig:pm_schematic}. 
Assuming that $t_{\rm ex}/\Sigma$ follows Eq.\ \eqref{eq:tex_asymm} , $x_\pm$ correspond to the locations where $d \ln \Sigma/d \ln |x| \approx 3$, and furthermore, the excited torques, $T_\pm$, follow from Eq.\ \eqref{eq:tpm_model} once the four numbers are known. 
Figure \ref{fig:tpm_model} compares this prediction for $T_\pm$ with what is found in the simulations, 
where the ``prediction'' makes use of the values of $x_\pm$ and $\Sigma_\pm$ extracted
from the simulations.  Above $K \sim 100$, the agreement is quite good, particularly at $K\gtrsim 3,000$, thus confirming
our claim that given $\Sigma$, one can calculate the one-sided torques.

Second, as we now show, $\Sigma_-$ may be determined by
integrating the  VSS equation from
$r=0$ to $r=r_p+x_-$. The calculation is nearly the
same as for the moderate gap case (cf.  the discussion surrounding Eq.\ \ref{eq:sm1}). Starting from 
Eq.\ (\ref{eq:far_inner}), but taking the upper limit of the integral to be $r_p+x_-$ yields
\be \label{eq:sm_high}
 {\Sigma_-\over\Sigma_{Z,-}} \approx  \frac{1}{1 + 0.27 K_-} , \label{eq:sigminus}
 \ee
where $\Sigma_{Z,-}$ is $\Sigma_Z$ at $x_-$ and where\footnote{Note that the asymmetric coefficient here is $1.26$ as opposed to $2.26$ because there is a factor of $r$ when converting between $F_\nu$ and $\Sigma$ for Keplerian disks.}
\be
K_- \equiv \frac{q^2}{\alpha h^2 |x_-|^3} \left(1 + 1.26 x_- \right) \label{eq:kminus}  \ .
\ee
Equation (\ref{eq:sigminus}) is the extension of Eq.\ (\ref{eq:sigsig}) to very deep gaps, but now
it only provides
a consistency relation between $\Sigma_-$ and $x_-$. 
In Figure \ref{fig:sm_model}, we show that the measured values of $\Sigma_-$ agree well with the values provided by Eq. \eqref{eq:sm_high} given $x_-$, for simulations with $K > 100$.

 We have now reduced the problem to three unknowns, $x_-$, $x_+$, and $\Sigma_+$. 
 One may additionally integrate the VSS equation between $x_-$ and $x_+$ to relate
 the jump in surface density 
 ($\Sigma_+-\Sigma_-$) to the torque deposited between $x_-$ and $x_+$.  In other words, 
 if one can determine the torque deposited between $x_-$ and $x_+$ (which is likely highly non-local and non-linear), as well as the values of
 $x_-$ and $x_+$, one will have a complete theory.

\subsection{Models based on local deposition are inadequate, particularly at high $K$} \label{sec:kocsis}

 A number of previous papers
 \citep{2010PhRvD..82l3011L,2012MNRAS.427.2660K,2012MNRAS.427.2680K}
  have constructed 1D models for what we call VSS. However, these
  are based on the assumption that $t_{\rm ex}=t_{\rm dep}$, i.e., they ignore the fact that
  waves transport angular momentum from where they are excited (at $\sim x_\pm$) to where
  they are deposited.  While that might seem a minor point, it leads to extremely erroneous results, 
  as we demonstrate briefly here. 
 
 To show this,
    we set $t_{\rm dep} = t_{\rm ex}$ in the VSS equation (Eq.\ \ref{eq:tdep}), and use Eq.\ \eqref{eq:tex_asymm} for $t_{\rm ex}$.
For simplicity, we also set $\dot{M} = 0$ and $\nu \ell = {\rm const}$ such that $F_\nu = 3 \pi \alpha h^2 \Sigma$. 
These approximations are purely for demonstration purposes since the full Eq.\ \eqref{eq:tdep} yields very similar results.
 It is straightforward to show that the solution of the VSS equation in the inner disk yields $\Sigma =  e^{-f_- K (h/|x|)^3}$, where we have set $\Sigma$ at the inner boundary to one and $f_- \approx 0.08$ from Eq.\ \eqref{eq:tex_asymm} evaluated at $x=-h$.
 Similarly, in the outer disk $\Sigma = \Sigma_p e^{ f_+ K (h/x)^3}$, where $f_+ \approx 0.1$. 
 To connect the inner and outer disk we assume that $F_\nu = {\rm const}$ between $x=-h$
 and $x=+h$ \citep[as was done in e.g.,][]{2012MNRAS.427.2660K,2012MNRAS.427.2680K}. 
 The total torque for the local model is then $\Delta T \approx e^{(f_+ - f_-) K} - 1$. 
We see that the gap becomes exponentially deep and the torque exponentially large for $K \gtrsim 50$. 
 We find this same divergence of $\Delta T$ in the full Eq.\ \eqref{eq:tdep} with $\dot{M} \neq 0$ and retaining all $r$ dependencies \citep[a similar result is found in][]{2010PhRvD..82l3011L}. 
 Such a divergence is incorrect.
 For example, at $K\sim 10^4$, the local model would predict $\Delta T/\dot{M}\ell_p\sim e^{200}$ for the pileup factor, 
 whereas we find a value $\Delta T/\dot{M}\ell_p\sim 10$, an enormous discrepancy.
We may conclude that local deposition is grossly inadequate, particularly at large $K$\footnote{We note that \citet{2012MNRAS.427.2660K,2012MNRAS.427.2680K} do not find such large $\Delta T$ values because they both reduce the coefficient of $t_{\rm ex}$, as well as include radially dependent $\alpha$ and $h$ profiles. }.

\section{Summary}\label{sec:conclusions}

We examined the planet-disk interaction problem in disks of low enough mass that the planet's migration time is slower than the disk's
viscous accretion time. Our main results are as follows:
\begin{itemize}
	\item One may study such disks by treating the planet's orbit as fixed, and examining the disk's properties
	in viscous steady state (VSS). This is a particularly clean setup to study the planet-disk
	interaction problem.  The key question becomes what is the total torque injected by the planet
	($\Delta T$) in VSS, for a particular set of problem parameters (principally, $q,\alpha,h$)?
	The value of $\Delta T$ determines both the pileup of disk material exterior to the planet's orbit,
	and the migration rate of the planet. 

	\item We predicted $\Delta T$ for moderately deep gaps (\S \ref{sec:moderate}). 
	We then ran a series of hydrodynamical simulations that reached VSS for a variety of parameters.
	The results of the simulations agreed with the theory for moderately deep gaps. But for very 
	deep gaps, the theory is inadequate. Empirically, for very deep gaps our simulations yielded the approximate
	relation $\Delta T/\dot{M}\ell_p \approx 4 (q/\alpha)$ when $q \gtrsim 10^{-4}$.  
 
	\item We calculated the resulting planet migration rate, showing how the well-understood
	Type I rate smoothly transitions into a new Type II rate as the gap formed by the planet becomes
	increasingly deep. 

\end{itemize}

\section{ Open Questions}
\label{sec:open}
We have left a considerable number of open questions to future investigations. Some of these are as follows.

\begin{itemize}

\item
What is the VSS result for parameter values not examined
in this paper, and is it possible  to achieve a pileup factor larger than the largest we found 
in our simulations (i.e., $\Delta T/\dot{M}\ell_p\sim 10$ for $K\sim 10^4$)?
Our simulations  only explored a limited range of parameters: we set $h=0.05$ 
and $q$ and $\alpha$ along the grid of filled circles in Figure \ref{fig:param}.  
We expect that both higher $q$ and lower $\alpha$ might lead to a higher pileup factor. 
At higher $q$, we found that the disk transitioned to an eccentric state (see also \citealt{2006MNRAS.368.1123G,2006A&A...447..369K,2008A&A...487..671K,2014ApJ...782...88F,2017MNRAS.467.4577T}). 
How realistic is that result, and if it is realistic, what is the VSS for an eccentric disk? 
A potential difficulty is that eccentric disks behave quite differently in 2D and 3D 
 \citep[e.g.,][]{2008MNRAS.388.1372O,2019ApJ...882L..11L}. 
Regarding lower $\alpha$,   we have not been able to reach VSS for $\alpha<10^{-4}$ with our simulations because of their
computational   cost.  Of course, if $\alpha$ is too small, the time to reach VSS might be longer than the
age of the disk.

\item 
Do 3D effects significantly affect the pileup?  How important is accretion onto the planet?
What is the effect of using a more realistic equation of state? \citet{2019ApJ...878L...9M}
show that an adiabatic (rather than locally isothermal) equation of state leads to a different $t_{\rm dep}$ profile.
We suspect the change will be minor because $\Delta T$ is  most sensitive to what happens very close to the planet, where the effect of equation of state is likely minor.

\item
 Are the surface density profiles for disks in VSS consistent with those inferred from observations
of protoplanetary disks? For example, could inferred inner holes be the result of the deficit of material
within a planet's orbit relative to a pileup outside of it? 
And could some of the gaps and rings imaged at large radii---that are often attributed to planets
\citep[e.g.,][]{2018ApJ...869L..47Z}---be the result of a pileup outside of the planet?
In this paper, we have only addressed gas dynamics, and to make detailed comparison with
observations one must also understand how the dust behaves. 
Hence we leave comparison with observations to future work.

\item How wide are the inner gaps? 
We find that for many of our highest $K$ simulations, the gap extends into our inner wave-killing zone. 
Future work should extend the inner boundary to smaller radii to determine  more realistic wave deposition profile. 
This may prove useful for diagnosing whether an observed gap is due to a planet or by some other process.

\end{itemize}

\acknowledgements
We thank the referee, Roman Rafikov, for a thorough reading of the manuscript and many helpful comments, and Diego Mu{\~n}oz for many insightful discussions.
This research was supported in part through the computational resources and staff contributions provided for the Quest high performance computing facility at Northwestern University which is jointly supported by the Office of the Provost, the Office for Research, and Northwestern University Information Technology. 
 Y.L. acknowledges NASA grant  NNX14AD21G and NSF grant AST-1352369.

\begin{appendix}
\section{Steady-state derivation} \label{sec:app_deriv}

We derive the equations for the three angular momentum densities (total, wave, and mean flow) that are needed for  \S \ref{sec:exdep}.
The 2D equations of motion for a fluid with surface density $\Sigma$, velocity $\mathbf{v}$, and pressure $P$ are,
\begin{align}
\ppderiv{\Sigma}{t} &+ \div \left( \Sigma \vec{v} \right) = 0 , \label{eq:app_2d_sig}\\
\ppderiv{\vec{v}}{t} &+ \vec{v} \cdot \del \vec{v} = - \del \Phi  - \frac{\del P}{\Sigma} -\frac{1}{\Sigma} \div (\nu \Sigma \vec{S})  , \label{eq:app_2d_mom}
\end{align}
where $\Phi$ is the external gravitational field and $\vec{S} = \del \vec{v}  + \del \vec{v}^T - 2/3 (\del \cdot \vec{v})$ is the stress tensor. 
Specializing to cylindrical coordinates, $(r,\phi)$, $\Sigma$ and the specific angular momentum, $\ell = r v_\phi$ evolve according to,
\begin{align}
\ppderiv{\Sigma}{t} &+ \frac{1}{r} \pderiv{r} \left(  r \Sigma v_r \right) + \frac{1}{r^2} \pderiv{\phi} \left(\Sigma \ell \right) = 0 , \label{eq:app_sigma} \\ 
\ppderiv{\ell}{t} &+ v_r \ppderiv{\ell}{r} + \frac{1}{2 r^2} \ppderiv{\ell^2}{\phi} = -\ppderiv{\Phi}{\phi} + \frac{1}{r \Sigma} \pderiv{r} \left( r^2 \nu \Sigma S_{r \phi} \right)   + \frac{1}{\Sigma} \ppderiv{\mathnormal{f}}{\phi} ,\label{eq:app_l} 
\end{align}
where $S_{r \phi} = r \partial_r \Omega + r^{-1} \partial_\phi v_r $ and, for convenience, we have combined the pressure and viscous stress into $\mathnormal{f}= - P + \nu \Sigma S_{\phi\phi}$. 
Together, Eqs.\ \eqref{eq:app_sigma} and \eqref{eq:app_l} describe the evolution of the total angular momentum density,
\begin{equation} \label{eq:app_ltot}
\pderiv{t} \left(\Sigma \ell \right) + \frac{1}{r} \pderiv{r} \left( r \Sigma v_r \ell - r^2 \nu \Sigma S_{r \phi} \right) + \frac{1}{r^2} \pderiv{\phi} \left( \Sigma \ell^2 \right) =  - \Sigma \ppderiv{\Phi}{\phi}  + \ppderiv{\mathnormal{f}}{\phi} ,
\end{equation}
and have the azimuthal averages,
\begin{align} 
2 \pi r  \pderiv{t} \avg{\Sigma} & - \ppderiv{\dot{M}}{r} = 0 , \label{eq:mdot_avg} \\
 2\pi r \pderiv{t}  \avg{\Sigma \ell} &+ \pderiv{r} \left( 2 \pi r \avg{ \Sigma v_r \ell}  + F_\nu \right) = t_{\rm ex}  , \label{eq:ltot_avg}
\end{align}
where $\dot{M}$ is defined in Eq.\ \eqref{eq:mdot};
$F_\nu \equiv - 2 \pi r^2 \avg{\nu \Sigma S_{r \phi} } $, as displayed in Eq.\ \eqref{eq:fnux}; 
and the excited torque density $t_{\rm ex}$ is defined in Eq.\ \eqref{eq:exex}.
To obtain the evolution equations for the wave angular momentum $\avg{\Sigma' \ell'}$, we add $\ell' \partial_t \Sigma$ to $\Sigma' \partial_t \ell$  and take the azimuthal average, 
\be \label{eq:wave_ang2}
   2 \pi r \pderiv{t} \avg{\Sigma' \ell'} + \pderiv{r} \avg{2 \pi r^2 \Sigma v_r v_\phi'} = t_{\rm ex} - t_{\rm dep} .
\ee
The term inside the radial derivative is the wave flux of angular momentum, defined in Eq.\ \eqref{eq:fwavex}
and $t_{\rm dep}$ is the deposition rate of angular momentum by the waves, 
\begin{align} \label{eq:tdep_def}
        \frac{1}{2 \pi r} t_{\rm dep}  =  \avg{ \Sigma' v_r'} \ppderiv{\avg{\ell}}{r}  
    &- \avg{\Sigma} \avg{ v_r' \ppderiv{\ell'}{r}} - \avg{\frac{\Sigma'}{\Sigma} \ppderiv{f}{\phi}}
    - \avg{ \frac{\Sigma'}{\Sigma} \pderiv{r} \left( r^2 \nu \Sigma S_{r\phi} \right)} .
\end{align}
Note that since $t_{\rm dep}$ depends on wave quantities, it should reach a steady-state value on the timescale for the waves to reach steady-state. 
Finally, to obtain the evolution of the axisymmetric angular momentum we subtract Eq.\ \eqref{eq:wave_ang2} from Eq.\ \eqref{eq:ltot_avg}, which results in Eq.\ \eqref{eq:lav}.

\subsection{Approximations} \label{sec:app_approx}

\begin{figure}
	\centering
	\includegraphics[trim={.3cm .3cm .2cm 0},clip,width=.88\textwidth]{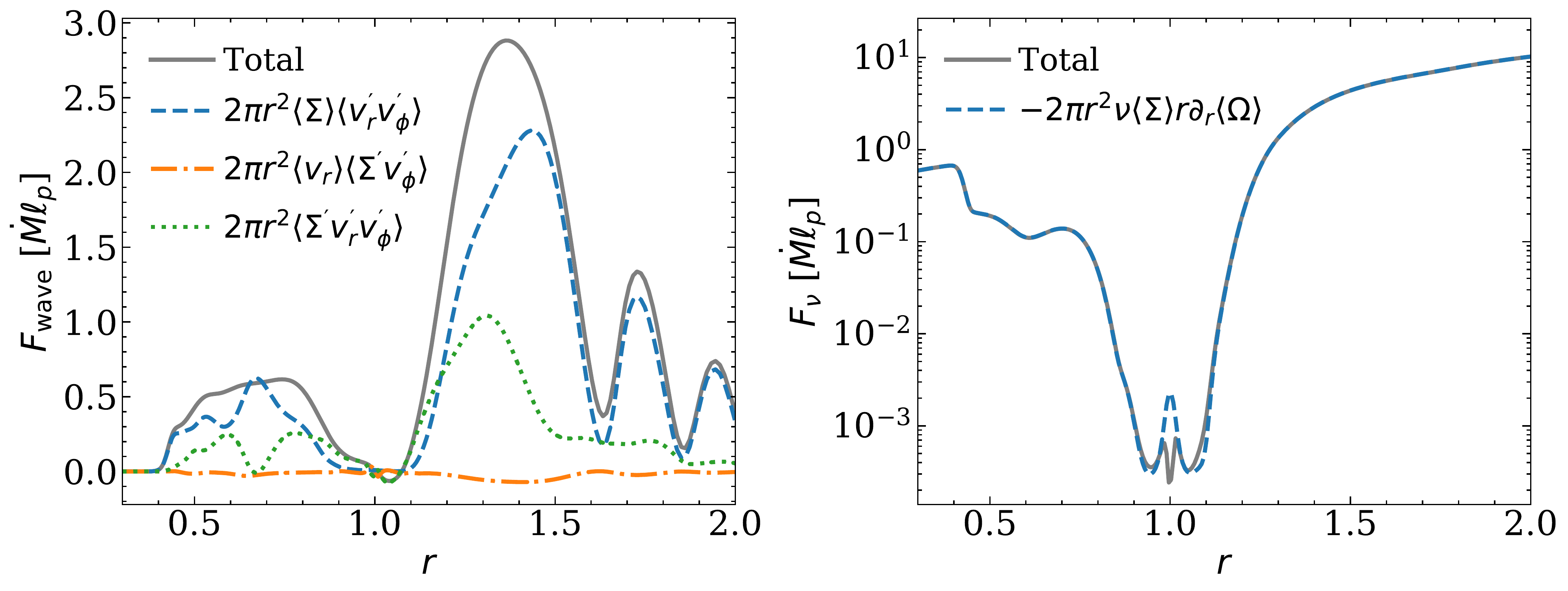}
	\caption{Radial profiles of $F_{\rm wave}$ and $F_{\nu}$ from an example hydrodynamical simulation. Left : Radial profiles of $F_{\rm wave}$ (grey line) and the three components defined in Eq.\ \eqref{eq:fwavex3}. The standard wave flux term $\propto \avg{v_r' v_\phi'}$ is shown as the blue dashed line while the triple correlation non-linear term is shown as the dotted green line. 
	The $\avg{v_r}$ term (orange dot-dashed) is negligible everywhere. 
	Right: The radial $F_\nu$ profile with (grey) and without (blue dashed) the $m>0$ components.}
	\label{fig:approx}
\end{figure}

The wave flux in Eq.\ \eqref{eq:fwavex} is made of three terms defined in Eq.\ \eqref{eq:fwavex3}.
In the left panel of Figure \ref{fig:approx}, we  show the total $F_{\rm wave}$ from an example simulation (described in \S\ref{sec:results}; solid line) and each of its terms.
The $\propto \avg{v_r' v_\phi'}$ term (dashed line) is dominant nearly everywhere in the disk except for the region closest to the planet where the triple correlation term (dotted line) becomes dominant. 
As expected, the $\propto \avg{v_r}$ term (dashed-dotted line) is nearly zero throughout the disk.

The viscous flux in Eq.\ \eqref{eq:fnux} is made of two terms,
\be
	F_\nu = -2 \pi r^2 \nu \avg{\Sigma} r \ppderiv{ \avg{\Omega} }{r}- 2 \pi r^2 \nu \avg{ \Sigma' \left(r \ppderiv{\Omega'}{r}  + \frac{1}{r} \ppderiv{v_r'}{\phi} \right)} ,
\ee
where the second term depends only on wave quantities. 
We show in the right panel of Figure \ref{fig:approx} that this wave term is negligible everywhere in the disk except for very close to the planet.

\section{Linear solution} \label{sec:app_linear}
We describe how we solve the linear response of a disk to a planet, which is needed in \S\ref{sec:dt_theory} and \S\ref{sec:excitation}.
We decompose variables as follows:
\begin{align}
\Sigma &= \avg{\Sigma} \left[ 1 + \Re \left\{ \sum_m \sigma_m e^{i m (\varphi - \Omega_p t)} \right\} \right] ,\\
 v_\phi &= r \avg{\Omega}  + \Re \left\{ \sum_m v_m e^{ i m (\varphi - \Omega_p t)} \right\} ,\\
 v_r &= \Re \left\{ \sum_m u_m e^{ i m (\varphi - \Omega_p t)} \right\} ,
\end{align}
where we have neglected the azimuthal average of the radial velocity (see Figure \ref{fig:approx}). 
The gravitational potential is similarly transformed to $\phi_m$. 
To obtain the linear equations of motion we expand Eqs.\ \eqref{eq:app_2d_sig} and \eqref{eq:app_2d_mom} with $P = c_s^2(r) \Sigma$ to first order in $(u_m,v_m,\sigma_m)$.
The result is \citep{1979ApJ...233..857G,1993Icar..102..150K,2002ApJ...565.1257T}
\begin{align}
   i m \left(\avg{\Omega} - \Omega_p \right) u_m &- 2 \Omega v_m + c_s^2\frac{d \sigma_m}{dr}  - \nu f_r = -\frac{d \phi_m}{d r} , \label{eq:lin_u} \\ 
    i  m \left( \avg{\Omega} - \Omega_p \right)v_m &+ \left(\frac{ \kappa^2}{2 \avg{\Omega}} \right) u_m + \frac{i m c_s^2 \sigma_m}{r} - \nu f_\phi =- \frac{i m \phi_m}{r} , \label{eq:lin_v} \\
    i m \left( \avg{\Omega} - \Omega_p \right) \sigma_m &+ u_m \frac{d \ln \avg{\Sigma}}{d r} + \frac{1}{r} \frac{d}{dr} \left( r  u_m\right) + \frac{i m  v_m}{r} = 0  ,\label{eq:lin_s}
\end{align}
where, $\kappa^2 = 4 \avg{\Omega}^2 + r d \avg{\Omega}^2/dr$. 
The viscous accelerations, $f_{r,\phi}$, are determined by linearly expanding $ (\del \cdot (\nu \Sigma \vec{S}))/\Sigma$ in Eq.\ \eqref{eq:app_2d_mom}.
Once $\avg{\Sigma}$ and $\avg{\Omega}$ are specified, we solve Eqs.\ \eqref{eq:lin_u}-\eqref{eq:lin_s} as a boundary value problem for each $m$. 
Since the equations are linear, we use a simple matrix method where we discretize the equations onto a radial grid, specify outgoing wave boundary conditions at both boundaries, and solve the resulting tri-diagonal system of equations. 
For the outgoing wave boundary conditions, we assume that far from the planet the waves are in the WKB limit and follow,
\begin{equation}
 \left\{u_m,v_m,\sigma_m\right\} \propto e^{i k_m r} ,
\end{equation}
where $k_m$ is the positive root of the WKB dispersion relation, $c_s^2 k_m^2 = \kappa^2-  m^2 (\avg{\Omega} - \Omega_p)^2 $. 
This matrix inversion method is different than the shooting method typically used to solve the linear planet-disk equations of motion \citep[e.g.,][]{1993Icar..102..150K,2002ApJ...565.1257T,2012ApJ...747...24R,2012ApJ...758...33P}. 

\section{Numerical Appendix} \label{sec:app_num}

Here we describe the numerical setup of our hydrodynamical simulations.
FARGO3D solves Eqs.\ \eqref{eq:app_2d_sig} \& \eqref{eq:app_2d_mom} on a staggered mesh where the density lies at the center of the cell and the velocities lie at the edges of the cell in their respective directions.
For simulating accretion disks, FARGO3D uses the fast advection algorithm of its predecessor to significantly increase the CFL limited timestep by removing the dominant Keplerian azimuthal velocities \citep{2000A&AS..141..165M}.
Typically, the timestep is constrained by the radial sound crossing time in the inner cells.

Our boundary conditions described in \S\ref{sec:bcs} apply only for the azimuthally averaged density and velocities.
To ensure that there are no waves at the boundaries we adopt wave-killing zones \citep{2006MNRAS.370..529D}. 
In these regions, we artificially enhance wave damping, such that the waves vanish at the computational boundaries. 
In particular, at the end of each timestep we additionally evolve the radial velocity according to
\begin{equation} \label{eq:wkz}
    \frac{\partial v_r}{\partial t} = -\left(\frac{v_r - \avg{v_r}}{\tau} \right) R(r),
\end{equation}
where the local damping timescale $\tau = 1/(30 \Omega_K(r))$ and $R(r)$ is a quadratic function which is zero in the bulk of the domain, and rises to unity near the boundaries \citep{2006MNRAS.370..529D}. 
Our choice to damp only the radial velocity ensures that we conserve both mass and angular momentum in the wave-killing zones. 
This is in contrast to most of the other gap-opening studies which utilize wave-killing zones that additionally damp $\ell$ and $\Sigma$ \citep[e.g.,][]{2013ApJ...769...41D,2015A&A...574A..52D,2017PASJ...69...97K}.

\subsection{Flux and torque calculation} \label{sec:app_compute}

Here we outline our numerical calculation of $\dot{M}$, $F_\nu$, $F_{\rm wave}$, $t_{\rm ex}$,  presented in the main text. 
In short, these quantities are taken from their respective steps in the FARGO3D algorithm and a running time average is computed at each timestep as FARGO3D evolves the equations of motion over the averaging periods given in Table \ref{tab:sims}.
As an example of this process, we focus here on calculating $F_{\rm wave}$ (Eq.\ \ref{eq:fwavex}) as it is the most involved. 
At each timestep, FARGO3D updates the angular momentum of a cell from the angular momentum fluxes in the $r$ direction. 
These fluxes are computed by reconstructing the cell-centered angular momenta to the radial faces of each cell \citep[for details of this process see][]{2016ApJS..223...11B}.
During these updates we store the  reconstructed values of $\Sigma$ and $\ell$ on the cell faces, as well as the $v_r$ values.
Using these we compute, 
\be \label{eq:fwavea}
F_{\rm wave}(r) = 2 \pi r \Re\left\{ \sum_{m>0} (v_r \Sigma^*)_m^\dagger  \ell_m^*  \right\}  ,
\ee
where $(v_r \Sigma^*)_m$ and $\ell^*$ are the $m$-th components of the Fourier transforms of $v_r \Sigma^*$  and $\ell^*$, and where the stars indicate that these values are reconstructed. 
Note that in the sum we only retain the $m > 0$ terms. 
This procedure is done every timestep during the averaging period, with each timestep contributing a new value to the running average for each $m$ contribution to $F_{\rm wave}$. 
An analogous procedure is done for $\dot{M}$ and $F_\nu$ in the update functions for $\Sigma$ and $v_\phi$, respectively.
We choose to calculate $F_{\rm wave}$, $F_\nu$, and $\dot{M}$ in this way so that (i) the time-averages exactly correspond to the changes in total angular momentum and mass in a given cell over the averaging period, and (ii) so that we may separate the contributions from different $m$ values.

\section{Simulation Table} \label{sec:app_table}

\definecolor{mya0}{rgb}{0.9098,0.0000,0.0431}
\definecolor{mya1}{rgb}{0.0000,0.2471,1.0000}
\definecolor{mya2}{rgb}{1.0000,0.7686,0.0000}
\definecolor{mya3}{rgb}{0.5412,0.1686,0.8863}
\definecolor{mya4}{rgb}{0.0118,0.9294,0.2275}

\begin{table*}
\label{tab:sims}
\centering
\caption{Overview of all simulations used in this study, grouped by $K$. For each simulation, the $\Delta T$ value in parenthesis corresponds to the result of the low resolution simulation. The quantities $\Sigma_{Z,\pm}$ refer to $\Sigma_Z$ at $x_\pm$.}
\def\arraystretch{1.5}
    \begin{tabular}{lcccccccccccc}
\hline
\hline
Name & $q$ & $\alpha$ & $K$ & $\Delta \dot{M}$ & $t_{\rm avg}$ & $x_-$ & $x_+$ & $\Sigma_-$ & $\Sigma_+$ & $\Delta T_{\rm LR}$ & $\Delta T_c$ & $\Delta T$\\
Units &  &  &  & $\%$ & $t_{\rm orb}$ & $r_p$ & $r_p$ & $\Sigma_{Z,-}$ & $\Sigma_{Z,+}$ & $\dot{M} \ell_p$ & $\dot{M} \ell_p$ & $\dot{M} \ell_p$\\
\hline
\textcolor{mya2}{q1x3a3x4} & 1e-03 & 3e-04 & 10667 & 9.8 & 1000 & -0.207 & 0.275 & 5.35e-02 & 0.719 & 15.8 & -4.84 & 11.0 (14.2)\\
\hline
\textcolor{mya0}{q3x4a1x4} & 3e-04 & 1e-04 & 2880 & 8.4 & 3000 & -0.166 & 0.198 & 7.17e-02 & 0.277 & 3.83 & -0.672 & 3.15 (4.11) \\
\textcolor{mya4}{q1x3a1x3} & 1e-03 & 1e-03 & 3200 & 1.2 & 3000 & -0.177 & 0.228 & 7.23e-02 & 0.276 & 2.58 & -0.581 & 2.00 (2.08)\\
\textcolor{mya1}{q2x3a3x3} & 2e-03 & 3e-03 & 4267 & 0.15 & 100 & -0.197 & 0.260 & 6.73e-02 & 0.239 & 1.99 & -0.733 & 1.25 (1.54)\\
\hline
\textcolor{mya2}{q3x4a3x4} & 3e-04 & 3e-04 & 960 & 0.80 & 3000 & -0.145 & 0.169 & 0.124 & 0.260 & 1.39 & -0.205 & 1.19 (1.34)\\
\textcolor{mya1}{q1x3a3x3} & 1e-03 & 3e-03 & 1067 & 0.66 & 200 & -0.150 & 0.191 & 0.108 & 0.218 & 0.913 & -0.324 & 0.590 (0.662)\\
\textcolor{mya3}{q2x3a1x2} & 2e-03 & 1e-02 & 1280 & 0.61 & 100 & -0.171 & 0.213 & 0.116 & 0.191 & 0.621 & -0.308 & 0.313 (0.385)\\
\hline
\textcolor{mya0}{q1x4a1x4} & 1e-04 & 1e-04 & 320 & 1.1 & 3000 & -0.118 & 0.126 & 0.148 & 0.272 & 0.963 & -5.76e-02 & 0.905 (0.830)\\
\textcolor{mya4}{q3x4a1x3} & 3e-04 & 1e-03 & 288 & 0.37 & 100 & -0.112 & 0.133 & 0.173 & 0.258 & 0.596 & -0.148 & 0.448 (0.459)\\
\textcolor{mya3}{q1x3a1x2} & 1e-03 & 1e-02 & 320 & 0.14 & 100 & -0.107 & 0.119 & 0.170 & 0.166 & 0.486 & -0.307 & 0.179 (0.213)\\
\hline
\textcolor{mya2}{q1x4a3x4} & 1e-04 & 3e-04 & 107 & 1.8 & 100 & -6.70e-02 & 8.41e-02 & 0.205 & 0.275 & 0.457 & -3.48e-02 & 0.422 (0.412)\\
\textcolor{mya1}{q3x4a3x3} & 3e-04 & 3e-03 & 96 & 1.1 & 100 & -7.28e-02 & 7.73e-02 & 0.281 & 0.278 & 0.340 & -0.144 & 0.196 (0.203)\\
\hline
\textcolor{mya0}{q3x5a1x4} & 3e-05 & 1e-04 & 29 & 6.7 & 10000 & -6.70e-02 & 7.06e-02 & 0.458 & 0.516 & 0.228 & -6.07e-03 & 0.222 (0.239)\\
\textcolor{mya4}{q1x4a1x3} & 1e-04 & 1e-03 & 32 & 0.51 & 100 & -6.70e-02 & 6.39e-02 & 0.461 & 0.473 & 0.219 & -3.76e-02 & 0.182 (0.188)\\
\textcolor{mya3}{q3x4a1x2} & 3e-04 & 1e-02 & 29 & 0.23 & 100 & -6.70e-02 & 6.39e-02 & 0.570 & 0.546 & 0.166 & -9.95e-02 & 6.62e-02 (6.74e-02)\\
\hline
\textcolor{mya2}{q3x5a3x4} & 3e-05 & 3e-04 & 9.6 & 1.7 & 100 & -6.70e-02 & 6.39e-02 & 0.733 & 0.748 & 0.108 & -1.23e-02 & 9.53e-02 (9.37e-02)\\
\textcolor{mya1}{q1x4a3x3} & 1e-04 & 3e-03 & 11 & 0.63 & 100 & -6.11e-02 & 5.73e-02 & 0.718 & 0.706 & 0.101 & -3.90e-02 & 6.17e-02 (6.21e-02)\\
\hline
\textcolor{mya0}{q1x5a1x4} & 1e-05 & 1e-04 & 3.2 & 2.9 & 100 & -6.70e-02 & 6.39e-02 & 0.893 & 0.898 & 4.56e-02 & -5.96e-03 & 3.96e-02 (3.87e-02)\\
\textcolor{mya4}{q3x5a1x3} & 3e-05 & 1e-03 & 2.9 & 0.67 & 100 & -6.70e-02 & 6.39e-02 & 0.904 & 0.902 & 3.75e-02 & -1.22e-02 & 2.54e-02 (2.57e-02)\\
\textcolor{mya3}{q1x4a1x2} & 1e-04 & 1e-02 & 3.2 & 0.15 & 100 & -6.11e-02 & 5.73e-02 & 0.892 & 0.879 & 3.52e-02 & -2.40e-02 & 1.12e-02 (1.15e-02)\\
\hline
\textcolor{mya2}{q1x5a3x4} & 1e-05 & 3e-04 & 1.1 & 0.86 & 100 & -6.70e-02 & 6.39e-02 & 0.964 & 0.963 & 1.57e-02 & -4.97e-03 & 1.07e-02 (1.08e-02)\\
\textcolor{mya1}{q3x5a3x3} & 3e-05 & 3e-03 & 0.96 & 0.31 & 100 & -6.70e-02 & 6.39e-02 & 0.966 & 0.964 & 1.28e-02 & -5.74e-03 & 7.01e-03 (7.37e-03)\\
\hline
\textcolor{mya4}{q1x5a1x3} & 1e-05 & 1e-03 & 0.32 & 0.27 & 100 & -6.70e-02 & 6.39e-02 & 0.990 & 0.989 & 4.50e-03 & -1.88e-03 & 2.62e-03 (3.20e-03)\\
\textcolor{mya3}{q3x5a1x2} & 3e-05 & 1e-02 & 0.29 & 0.26 & 100 & -6.70e-02 & 6.39e-02 & 0.990 & 0.990 & 3.55e-03 & -1.31e-03 & 2.24e-03 (2.23e-03)\\
\hline
\textcolor{mya1}{q1x5a3x3} & 1e-05 & 3e-03 & 0.11 & 0.17 & 100 & -6.70e-02 & 6.39e-02 & 0.997 & 0.997 & 1.40e-03 & -4.84e-04 & 9.17e-04 (1.08e-03)\\
\hline
\textcolor{mya3}{q1x5a1x2} & 1e-05 & 1e-02 & 0.03 & 0.24 & 100 & -6.70e-02 & 6.39e-02 & 1.000 & 1.000 & 3.80e-04 & -8.20e-05 & 2.99e-04 (3.02e-04)\\
\hline

\end{tabular}
\label{tab:simulations}
\end{table*}

\section{Ward Torque} \label{sec:app_ward}

\begin{figure}
	\centering
	\includegraphics[trim={.3cm .3cm .2cm 0},clip,width=.48\textwidth]{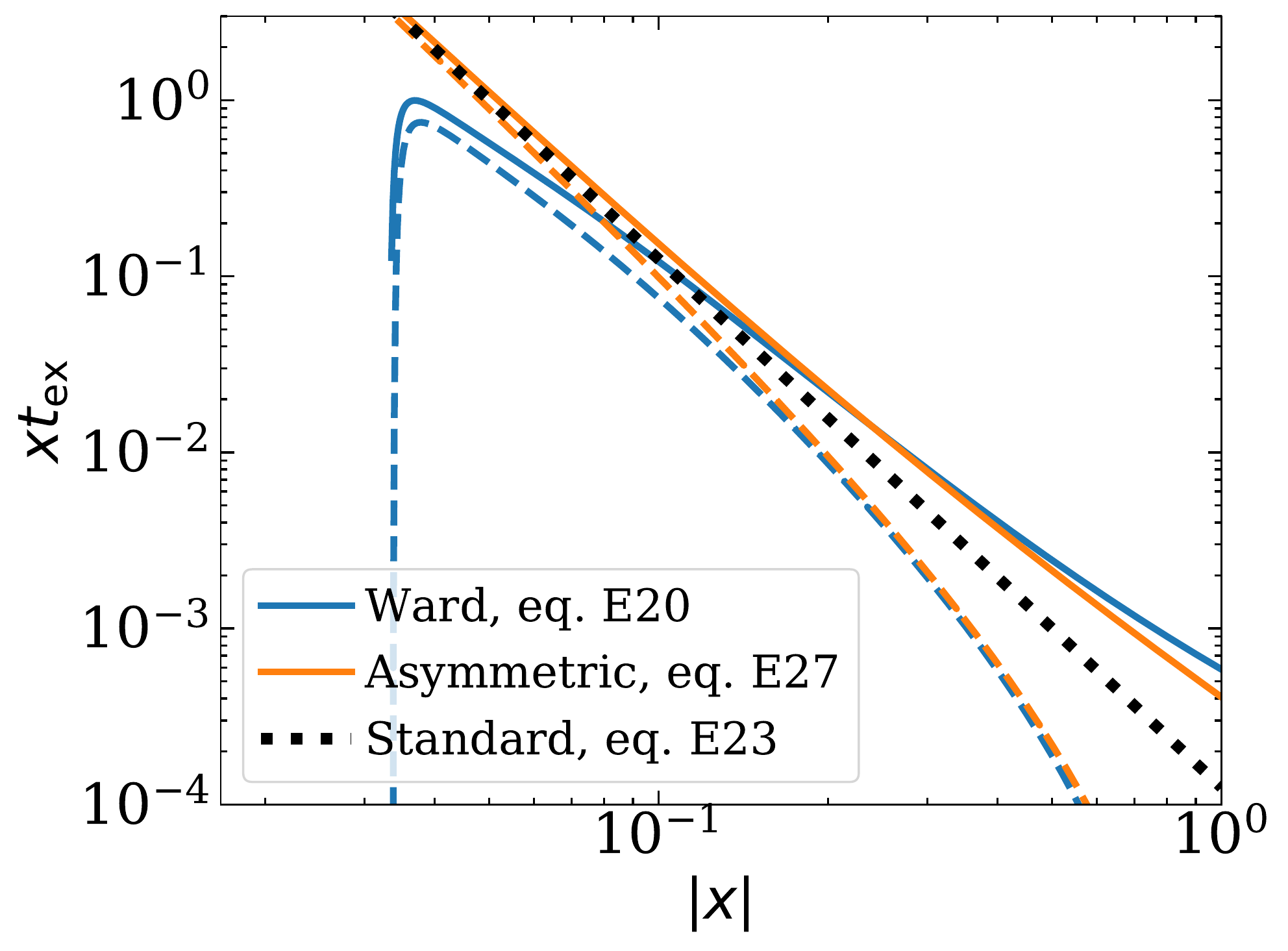}
	\caption{Comparison of different analytic $t_{\rm ex}$ profiles for $\Sigma = {\rm const}$. In the outer disk, Eq.\ \eqref{eq:app_tex_asymm} (orange solid) agrees with Ward's $t_{\rm ex}$ (blue solid; Eq.\ \eqref{eq:tex_ward0}) to within $3\%$ near $x \approx 0.2$, and both lie above the standard torque formula (black dotted; Eq.\ \eqref{eq:tex_gt80}. 
	Similarly, in the inner disk Eq.\ \eqref{eq:app_tex_asymm} agrees with Ward near $x \approx -0.2 $ to $10\%$, with both being below the standard torque formula.}
	\label{fig:torque_functions}
\end{figure}

We derive Eq.\ (\ref{eq:tex_asymm}), which is the leading order asymmetric correction to the standard torque formula
 \citep{1980ApJ...241..425G},  starting from the more general torque profiles of \citet{1993ApJ...419..155A} and \citet{1997Icar..126..261W}. 
\citet{1997Icar..126..261W} gives the excited torque density as (his Eq.\ 14, but see also Eq.\ 51 and 54 of  \citet{1993ApJ...419..155A}),

\begin{equation} \label{eq:tex_ward0}
      t_{\rm WW} =  \pm 2 q^2 \Sigma  \left(\frac{r}{r_p} \right)^2 \left(\frac{\Omega_K}{\kappa}\right)^2  m^4 \frac{\psi^2}{1 + 4 \xi^2}  r_p^3 \Omega_p^2 ,
\end{equation}
where $\xi = m c_s/(r \kappa)$, and the upper (lower) sign corresponds to the outer (inner) disk.
The potential $\psi$ is,
\begin{equation} \label{eq:psi}
    \psi = \frac{1}{2} \left(1 + \frac{r_p}{r} \right) K_1(\Lambda) + \left(2 m f + \frac{\epsilon}{2 m} \right)K_0(\Lambda) \sqrt{\frac{r_p}{r}} .
\end{equation}
Here, $K_0$ and $K_1$ are modified Bessel functions of the second kind,  $\Lambda = m |x/r_p|/\sqrt{(r/r_p)}$ and $f = |\Omega-\Omega_p|/\Omega_K$. 
This form of $\psi$ was derived in \citet{1997Icar..126..261W} (as opposed to \citet{1993ApJ...419..155A}), and so from now on we shall associate this particular form of the torque density to Ward. 
Specializing to a sound speed profile of $c_s = h r \Omega_K$ we can rewrite $\xi$ as $\xi = m h (\Omega_K/\kappa) \approx m h$ for a nearly Keplerian disk.
Equations \eqref{eq:tex_ward0} and \eqref{eq:psi} are evaluated at effective Lindblad resonances defined by $D_\star = \kappa^2 - m^2 (\Omega - \Omega_p)^2 + (m c_s/r)^2 = 0$. 
In terms of $m$ and $f$ this resonance condition is ,
\begin{equation} \label{eq:app_res}
    m^2 = \frac{1}{f^2 - h^2} \left(\frac{\kappa}{\Omega_K} \right)^2,
\end{equation}
i.e. for a given distance to the planet there is a corresponding value of $m$. 
Note that $m$ diverges to infinity as $f \rightarrow h$ as $|x| \rightarrow 2/3 h$, but the torque does not diverge due to the exponential decay of the Bessel functions with $\Lambda \propto m \rightarrow \infty$.

The \citet{1980ApJ...241..425G} approximation to the excited torque follows from setting  $r=r_p$ in Eqs.\ \eqref{eq:tex_ward0} and \eqref{eq:psi} unless it appears as $x = r-r_p$ in which case $|x| \approx 2 r_p/(3m)$. 
With these approximations $\psi = 2 K_0(2/3) + K_1(2/3)$ and the torque density becomes,

\be \label{eq:tex_gt80}
	t_{\rm GT} =   \pm \mathcal{C} \Sigma q^2 \left( \frac{r_p}{x} \right)^4 r_p^3 \Omega_p^2 ,
\ee  
where the numerical constant $\mathcal{C} = (32/81) (2 K_0(2/3) + K_1(2/3))^2 \approx 2.5$.
This is the $t_{\rm ex}$ profile of \citet{1980ApJ...241..425G}. 
Note that this is symmetric with respect to the sign of $x$. 
\citet{1997Icar..126..261W} showed that the leading order correction to the  \citet{1980ApJ...241..425G} torque follows from the $m^4 \psi^2$ term in Eq.\ \eqref{eq:tex_ward0},
\be
m^4 \psi^2 \approx \psi_0^2 \left(m^4 \pm 0.84  m^3 \right) ,
\ee
where $\psi_0 = 2 K_0(2/3) + K_1(2/3)$.
To convert $m$ to $x$ we expand the resonance condition (Eq.\ \ref{eq:app_res}) to $(r_p/|x|)^3$ order,

\begin{align}
m^4 \approx& \left(\frac{2}{3} \right)^4 \left[ \left(\frac{r_p}{x}\right)^4 \mp \left( \frac{r_p}{|x|} \right)^3 \right] , \\
m^3 \approx& \left( \frac{2}{3} \right)^3 \left( \frac{r_p}{|x|} \right)^3 .
\end{align}
The final torque density with the leading asymmetry is then,

\be\label{eq:app_tex_asymm}
t_{\rm ex} \approx t_{\rm GT} \left(1 + 2.26 \frac{x}{r_p} \right) .
\ee

In Figure \ref{fig:torque_functions}, we compare the full \citet{1997Icar..126..261W} torque given by Eq.\ \eqref{eq:tex_ward0} as a function of $x$ for a constant surface density disk against the \citet{1980ApJ...241..425G} approximation given by Eq.\ \eqref{eq:tex_gt80}. 
There are a few important points to highlight here. 
First, there is no torque asymmetry due to $t_{\rm GT}$. 
Second, far from the torque cutoff, Eq.\ \eqref{eq:app_tex_asymm} is a good approximation to $t_{\rm WW}$.

\label{lastpage}
\end{appendix}

\end{document}